\documentclass[USenglish,oneside,twocolumn]{article}

\usepackage[utf8]{inputenc}
\usepackage[big]{dgruyter_NEW}
\usepackage{chngcntr}
\usepackage{xspace}
\usepackage[boxruled]{algorithm2e}
\usepackage{color,xcolor,colortbl}
\usepackage{amsmath,amssymb}
\usepackage{enumitem}
\usepackage{natbib}
\usepackage{subfigure}
\usepackage{environ}
\usepackage[subtle]{savetrees}
\usepackage[T1]{fontenc}
\usepackage{lmodern}
\usepackage{changepage}
\usepackage{hyperref}



\newif\ifanonymous{}
\newif\ifextended{}

\anonymousfalse{} 
\extendedtrue{}  

\graphicspath{{../graphics/}{.}}
\DeclareMathAlphabet{\mathcal}{OMS}{cmsy}{m}{n}

\newcommand{\squishlist}{
	\begin{list}{$\bullet$}
		{ \setlength{\itemsep}{0pt}      \setlength{\parsep}{3pt}
			\setlength{\topsep}{3pt}       \setlength{\partopsep}{0pt}
			\setlength{\leftmargin}{3.5mm} \setlength{\labelwidth}{1em}
			\setlength{\labelsep}{0.5em} } }
	
	\newcommand{\squishend}{
\end{list}  }

\theoremstyle{definition}
\newtheorem{theorem}{Property}
\SetAlgorithmName{Protocol}{protocol}{List of Protocols}

\newcommand{\prifi}{PriFi\xspace}
\newcommand{\server}{guard\xspace}
\newcommand{\Server}{Guard\xspace}
\newcommand{\servers}{guards\xspace}

\newcommand{\setup}{\texttt{Setup}\xspace}
\newcommand{\anonymize}{\texttt{Anonymize}\xspace}
\newcommand{\blame}{\texttt{Blame}\xspace}

\newcommand{\rpm}{\raisebox{.2ex}{$\scriptstyle\pm$}}
\newcommand{\sfsize}{\fontsize{0.72\baselineskip}{0.68\baselineskip}\selectfont}

\newcommand{\sansbf}[1]{\textsf{\textbf{\sfsize{}\mbox{#1}}}}
\newcommand{\para}[1]{\vs{0.48em} \noindent \sansbf{\mbox{#1}}}

\newcommand{\ie}{{\em i.e., \xspace\/}}
\newcommand{\eg}{{\em e.g., \xspace\/}}

\renewcommand{\vec}[1]{\underline{#1}}
\newcommand{\sig}[2]{\mathcal{S}_{#1}{(#2)}}
\newcommand{\enc}[2]{\mathcal{E}_{#1}{(#2)}}
\newcommand{\dec}[2]{\mathcal{D}_{#1}{(#2)}}
\newcommand{\bin}{{\{0, 1\}}}

\newcommand{\ad}{\mathcal{A}}

\definecolor{azure}{rgb}{0.0, 0.4, 0.9}
\newcommand{\hl}{\color{blue}}
\newcommand{\bk}{\color{black}}

\DOI{foobar}

\cclogo{\includegraphics{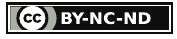}}

\begin{document}
	
	\ifanonymous{}
		\renewcommand{\prifi}{LLDC\xspace}
	\else
		\author*[1]{Ludovic Barman}
		\author[2]{Italo Dacosta}
		\author[3]{Mahdi Zamani}
		\author[4]{Ennan Zhai}
		\author[5]{Apostolos Pyrgelis}
		\author[6]{Bryan Ford}
        \author[7]{Joan Feigenbaum}
		\author[8]{Jean-Pierre Hubaux}
		
		\affil[1]{EPFL, E-mail: ludovic.barman@epfl.ch}	
		\affil[2]{UBS, E-mail: italo.dacosta@epfl.ch}
		\affil[3]{Visa Research, E-mail: mzamani@visa.com}
		\affil[4]{Alibaba Group, E-mail: ennan.zhai@alibaba-inc.com }
		\affil[5]{EPFL, E-mail: apostolos.pyrgelis@epfl.ch}
		\affil[6]{EPFL, E-mail: bryan.ford@epfl.ch}
        \affil[7]{Yale University, E-mail: joan.feigenbaum@yale.edu}
		\affil[8]{EPFL, E-mail: jean-pierre.hubaux@epfl.ch}
	\fi
	
	\title{\huge \prifi: Low-Latency Anonymity for \\Organizational Networks}
	\runningtitle{\prifi: Low-Latency Anonymity for Organizational Networks}
	
    \ifextended
        \newcommand{\vs}[1]{}
    \fi
    
	\begin{abstract}{
Organizational networks are vulnerable to traffic-analysis attacks that enable adversaries to infer sensitive information from network traffic --- even if encryption is used.
Typical anonymous communication networks are tailored to the Internet and are poorly suited for organizational networks.
We present \prifi, an anonymous communication protocol for LANs, which protects users against eavesdroppers and provides high-performance traffic-analysis resistance. 
\prifi builds on Dining Cryptographers networks (DC-nets), but reduces the high communication latency of prior designs via a new client/relay/server architecture, in which a client's packets remain on their usual network path without additional hops, and in which a set of remote servers assist the anonymization process without adding latency.
\prifi also solves the challenge of equivocation attacks, which are not addressed by related work, by encrypting traffic based on communication history.
Our evaluation shows that \prifi introduces modest latency overhead ($\approx100$ms for $100$ clients) and is compatible with delay-sensitive applications such as Voice-over-IP\@.
}
\end{abstract}

	
	\journalname{Proceedings on Privacy Enhancing Technologies}
    \DOI{Editor to enter DOI}
    \startpage{1}
    \received{..}
    \revised{..}
    \accepted{..}
    
    \journalyear{..}
    \journalvolume{..}
    \journalissue{..}
	
	\maketitle
	
	\vs{-1.4cm}

\section{Introduction}
\label{sec:intro}

{
	\begin{figure}[t]
		\centering
		\ifanonymous{}
		\includegraphics[width=23em]{prifi-model-v8-lldc.pdf}
		\else	
		\includegraphics[width=23em]{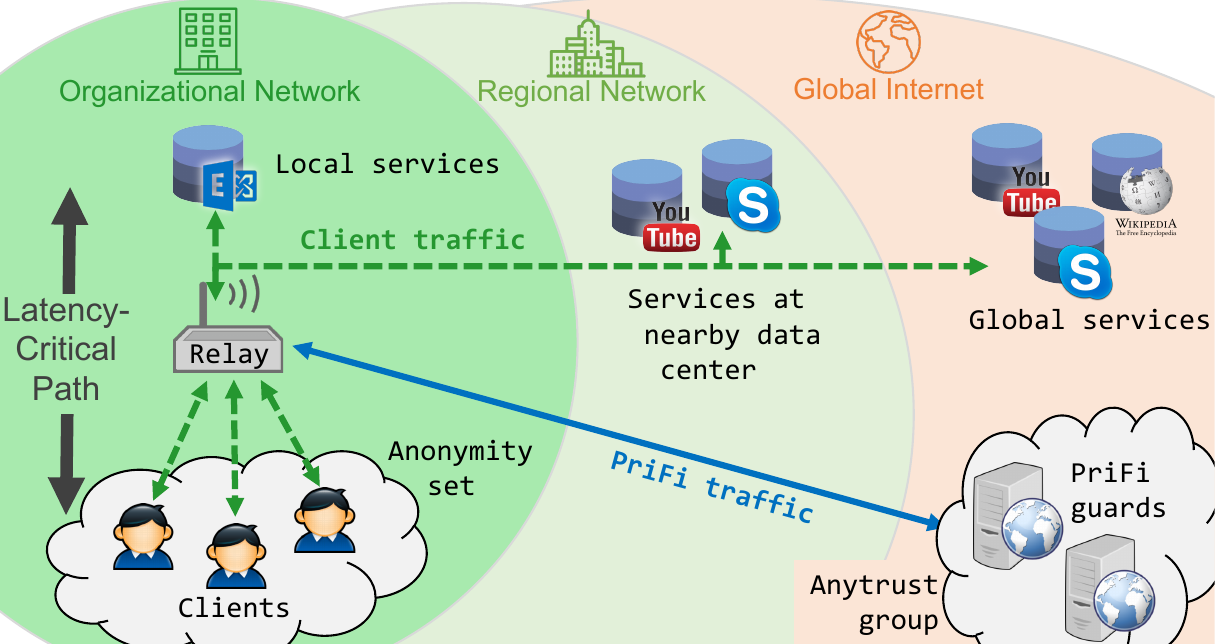}
		\fi
		\caption{\prifi's architecture consisting of clients, a relay, and a group of anytrust servers called the  \emph{\servers}. Clients' packets remain on their usual network path without additional hops, unlike in mix-networks and onion-routing protocols.}
		\label{fig:network_setup}
	\end{figure}
}


Local-area networks (LANs and WLANs) deployed in organizational networks are vulnerable to eavesdropping attacks.
Sensitive traffic is usually encrypted, but metadata such as \emph{who is communicating} and the communication patterns remain visible.
Such metadata enable an eavesdropper to identify and track users passively~\citep{franklin2006passive,xu2016device}, to infer contents and endpoints~\citep{wang2013improved,cai2012touching,gong2010fingerprinting,panchenko2011website,dubin2017know,schuster2017beauty,chang2008inferring,white2011phonotactic}, and potentially to perform targeted attacks on high-value devices and users.
Eavesdropping attacks can be performed by a single compromised endpoint or malicious user.
This is particularly worrisome when the users are loosely trusted, or when the organizational network is deployed in an adverse environment. For example, the International Committee of the Red Cross (ICRC) has strong privacy and security needs regarding their communications: a previous study confirms that its ``staff and beneficiaries need to communicate in a multitude of adverse environments that are often susceptible to eavesdropping, to physical attacks on the infrastructure, and to coercion of the personnel''~\citep{le2018enforcing}.


To protect against eavesdropping in LANs, few solutions exist today.
Anonymous communication networks (ACNs) are designed to conceal the communicating entities; however, most ACNs are designed for the Internet and translate poorly to the LAN setting.
Most ACNs rely on mix-networks or onion-routing, a common drawback of which is that their security relies on routing the traffic through a series of servers distributed around the Internet~\citep{dingledine2004tor,piotrowska2017loopix,chen2015hornet,chen2018taranet,kwon2017atom}.
First, this implies that internal communications in the organization's network would need to be routed over the Internet.
More importantly, to minimize the risks of coercion and collusion, these servers are typically spread across different jurisdictions, hence these designs introduce significant latency overhead.

Dining Cryptographers networks (DC-nets)~\citep{chaum1988dining} are an anonymization primitive that could be attractive in terms of latency in some contexts, as their security relies on information coding and not on sequential operations done by different servers.
In theory, therefore, anonymity can be achieved without high-latency server-to-server communication. This theoretical appeal has not been achieved in practice, however.  Previous DC-net systems such as Dissent~\citep{corrigan2010dissent}, Dissent in Numbers~\citep{wolinsky2012dissent}, and Verdict~\citep{corrigan2013verdict} still use costly server-to-server communication, thus imposing latencies in the order of seconds~\citep{wolinsky2012dissent}, and notably routing users' traffic through all of the servers.

We present \prifi, the first low-latency anonymous communication network tailored to organizational networks.
\prifi provides anonymity against global eavesdroppers: Users are assured that their communication patterns are indistinguishable from the communications of other \prifi users, even if the local network infrastructure is compromised.
To anonymize IP packet flows, \prifi works at the network level like a VPN\@.
Unlike a VPN, however, PriFi's security does not depend on a single endpoint, and the protocol provably resists traffic analysis.
\prifi provides low-latency, traffic-agnostic communication suitable for delay-sensitive applications such as streaming and VoIP, at the cost of higher LAN bandwidth usage.

Compared with previous work, \prifi significantly reduces communication latency through a new three-tier architecture composed of \emph{clients},
a \emph{relay} in the LAN (\eg a router), and \emph{\server} servers that are geographically distributed over the Internet (Figure~\ref{fig:network_setup}).
This architecture is compatible with organizational networks, and enables \prifi to avoid major latency overheads present in other ACNs.
Unlike previous DC-net systems that use multi-hop, multi-round protocols, and costly server-to-server communications~\citep{corrigan2010dissent, wolinsky2012dissent, corrigan2013verdict}, \prifi achieves similar guarantees while removing all server-to-server communications from the latency-critical path.
To produce anonymous output, \prifi ciphertexts pre-computed and sent by the \servers are combined locally at the relay, so that relay$\leftrightarrow$\server delay does not affect the latency experienced by clients.
Moreover, the traffic from clients remains on its usual network path, client$\leftrightarrow$relay$\leftrightarrow$destination, and does not go through the \servers. Some added latency results from buffering and software processing, but not from additional network hops. As a result, \prifi's latency is $2$ orders of magnitude lower than the closest related work with the same setup~\citep{wolinsky2012dissent}. 


We also present a solution for \emph{equivocation attacks}, \ie de-anonymization by a malicious relay sending different information to different clients and analyzing their subsequent behavior.
Previous DC-net systems are vulnerable to this attack, but do not address it~\citep{corrigan2010dissent, wolinsky2012dissent, corrigan2013verdict}.
Equivocation attacks can be detected using consensus or gossiping between the clients, at a high bandwidth and latency cost.  We present a new low-latency, low-bandwidth solution that relies on binding encryption to communication history.

We evaluate \prifi on a topology corresponding to an organizational network.
We observe that the latency overhead caused by \prifi is low enough for VoIP and video conferencing ($\approx100$~ms for $100$ users), and that the internal and external bandwidth usage of \prifi is acceptable in an organizational network ($\approx40$~Mbps in a $100$~Mbps LAN).
In comparison, the latency of the closest related work, Dissent in Numbers~\citep{wolinsky2012dissent}, is $14.5$~seconds for $100$ clients on the same setup.
One part of the evaluation is dedicated to the ICRC scenario; we replay real network traces recorded at an ICRC delegation and find that the increase in latency is tolerable in practice (between $20$ and $140$ms on average).

\vs{0.5em}
\noindent In this paper, we make the following contributions:

\squishlist
\item \prifi, a low-latency, traffic-agnostic, traffic-analysis-resistant anonymous communication network, building on a new DC-nets architecture optimized for LANs;
\item A low-latency method of protecting DC-nets against disruption attacks (\ie jamming) by malicious insiders;
\item A new low-latency defense against equivocation attacks;
\item An open-source implementation of \prifi, tested and evaluated on desktop computers, and implementations for Android and iOS~\citep{prifi-github};
\item An analysis of the effect of user mobility on DC-nets.
\squishend

    \section{Background on DC-nets}
\label{sec:background}

A Dining Cryptographers network or DC-net~\cite{chaum1988dining} is a protocol that provides anonymous broadcast for a group of users who communicate in lock-step, in successive rounds.
In a given round, each user produces a ciphertext of the same length.
One user also embeds a plaintext in its ciphertext.
Combining all users' ciphertexts reveals the plaintext, without revealing \emph{which} user sent it.

{
    \begin{figure}[t]
        \centering
        \includegraphics[width=15em]{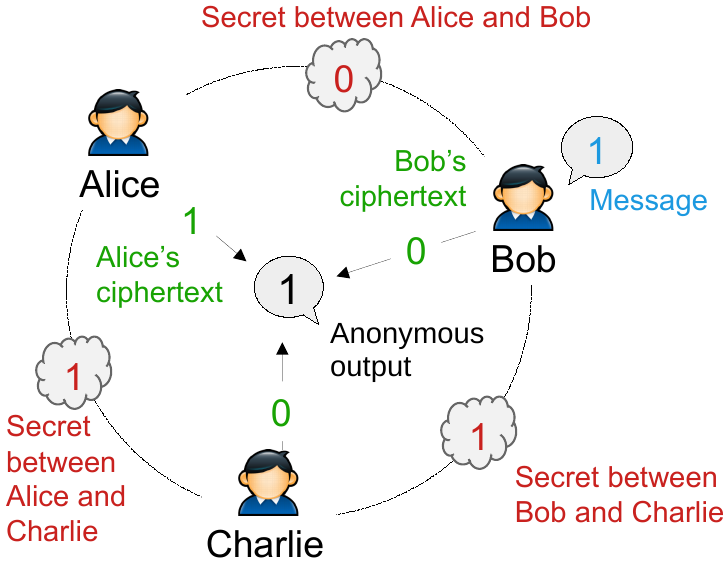}
        \caption{Example of a 1-bit DC-net.}
        \label{fig:dc-nets}
    \end{figure}
}

Figure~\ref{fig:dc-nets} shows an example of a 1-bit DC-net. Each pair of users derives a shared secret (in red).
Each user's ciphertext is the XOR of his shared secrets (in green).
Bob, the anonymous sender, also XORs in its message (in blue).
By XORing all ciphertexts together, all shared secrets cancel out, revealing the anonymized message.
If at least two users are honest, this protocol achieves unconditional sender anonymity. Every user sends a ciphertext of the same length, ciphertexts from honest parties are indistinguishable from each other without all shared secrets, and a single missing ciphertext (from an honest party) prevents the computation of the output.

In practice, to produce ciphertexts for multiple rounds, shared secrets are used to seed pseudo-random generators.

\para{Impact of Topology.} The original DC-net design~\cite{chaum1988dining} requires key material between every pair of members.
Dissent~\cite{wolinsky2012dissent} and subsequent work~\cite{corrigan2013verdict} present a more scalable two-tier topology made of clients and servers, which reduces the number of keys. 
However, the client/server topology has a significant negative impact on latency: It requires several server-to-server rounds of communication to ensure integrity, accountability, and to handle churn.
\prifi avoids this drawback with a new client/relay/\server architecture.

\para{Disruption Protection.} Vanilla DC-nets are vulnerable to disruption attacks from malicious insiders~\cite{chaum1988dining}, where a malicious user can corrupt other clients' messages.
Some previous work uses proactively verifiable constructions that are too slow for low-latency communication~\citep{corrigan2013verdict}. Others use ``trap bits'' and ``blame mechanisms''~\citep{wolinsky2012scalable} that require minutes to hours to find disruptors, mainly due to expensive server-to-server communication.
\prifi uses a new retroactive blame mechanism to detect disruptors in seconds.

\para{Equivocation Protection.} Answers to anonymous messages are typically broadcast to all users.
Previous DC-net designs did not address equivocation attacks,
where a malicious server sends each client different, identifiable information to distinguish the anonymous receiver (see Section~\ref{sec:protocol-equivocation}).
Equivocation thus represents a practical and covert attack vector against prior systems~\citep{corrigan2010dissent,wolinsky2012dissent,corrigan2013verdict}.
Equivocation attacks can be detected using consensus~\citep{lamport2001paxos} or gossiping~\citep{shah2009gossip} between the clients, at a high bandwidth and latency cost.
In \prifi, in contrast, messages from clients cannot be decrypted if an equivocation attack occurs, and \prifi  protects against this threat without communication among clients.



	\section{System Overview}
\label{sec:system-model}

\prifi is similar to a low-latency relay or gateway service within a LAN,
like a VPN or SOCKS proxy,
which tunnels traffic between clients and the relay (\eg a LAN router).
Informally, these tunnels protect honest clients' traffic from eavesdropping attacks.
The traffic is anonymized, preventing a third-party from assigning a packet or flow to a specific device or end-user.
Additionally, unlike traditional proxy services, (1) the relay need not be trusted, \ie security properties hold in case of compromise, and (2) the communications provably resist traffic-analysis attacks.

\subsection{System Model} 
\label{subsec:system-model}

Consider $n$ clients~$C_1,\dots,C_n$ that are part of an organizational network and are connected to a \emph{relay} $R$.
The relay is the gateway that connects the LAN to the Internet (\eg a LAN or WLAN router, Figure~\ref{fig:network_setup}) and typically is already part of the existing infrastructure.
The relay can process regular network traffic, in addition to running the \prifi software. Hence, \prifi can be deployed onto an existing network with minimal changes.

In the Internet, there is a small set $S_1,\dots,S_m$ of $m$ servers, called \emph{\servers}, whose role is to assist the relay in the anonymization process.
These \servers could be maintained by independent third parties, similar to Tor's volunteer relays, or sold as a ``privacy service'' by companies.
To maximize diversity and collective trustworthiness, these \servers are distributed around the world, preferably across different jurisdictions.
Therefore, the connections between the \servers and the relay are assumed to be high latency.

\subsection{Threat Model}
\label{subsec:threat-model}

Let $\ad$ be a computationally-bounded global passive adversary who observes all network traffic.
In addition to the passive adversary, as \prifi is a closed-membership system, we consider and address active attacks from insiders, but not active attacks from outsiders, which are fairly orthogonal to \prifi and can be addressed via adequate server provisioning~\citep{poletto2010architecture} or denial-of-service protection~\citep{peng2003protection}.

Most clients may be controlled by the adversary $\ad$,
but we require at least two honest clients at all times:
otherwise, de-anonymization is trivial.
The \servers are in the \emph{anytrust} model~\cite{corrigan2010dissent,wolinsky2012dissent,van2015vuvuzela}: we assume that least one \server is honest,
but a client does not need to know which one,
and we assume all \servers are highly available.

The relay is considered \emph{malicious but available}:
it may actively try to de-anonymize honest users or perform arbitrary attacks, but it will not perform actions that only affect the availability of \prifi communications such as delaying, corrupting, or dropping messages.
On one hand, a fully-malicious relay would make little sense in our setting, where it is the gateway that connects the LAN to the Internet: 
due to its position in the network, it can degrade or deny service for any protocol anyway (\eg drop all packets).
However, the relay is part of the infrastructure of the organization, and users would take administrative actions if the network is not operating properly.
On the other hand, an adversarial model where the relay is honest-but-curious would be too weak: In practice, if the relay is compromised (\eg it gets hacked), it could perform active attacks to de-anonymize clients.
Therefore, we use the ``malicious but available'' formulation to reflect that the relay needs to forward messages faithfully in order to provide service,
but if the relay maliciously attacks the protocol,
\emph{only} availability and not user privacy will suffer.

\subsection{Goals}
\label{subsec:system-model-goals}

\subsubsection{Security Goals}
\squishlist
\item \textbf{[G1] Anonymity}: An adversary has a negligible advantage in attributing an honest user's message to its author.
This includes traffic-analysis resistance:\eg the adversary can observe network-level traffic features.
\item \textbf{[G2] Accountability}: Misbehaving insiders are traceable without affecting the anonymity of honest users.\footnote{This definition of accountability should not be confused with other definitions
in which participants may de-anonymized based on communication content,
\ie if someone does not like what they say.}
\squishend

\vs{-0.5cm}
\subsubsection{System Goals}
\squishlist
\item \textbf{[G3] Low latency}: The delay introduced by \prifi should be small enough to support network applications with high QoS requirements,
\eg Voice-over-IP (VoIP) and videoconferencing applications.
\item \textbf{[G4] Scalability}:
One \prifi relay should support small to medium organizations
of up to a few hundred users,
a number typically observed in ICRC sites (Figure~\ref{fig:icrc-cdf}).
\squishend

\vs{-0.5cm}
\subsubsection{Non-Goals}
\label{subsec:system-model-nogoals}

\prifi does not target the following goals:

\squishlist
	\item Hiding \emph{all} traffic features: \prifi protects an honest user's traffic among all honest users' traffic, but does not hide global/aggregate communication volumes or time series of packets.
		Informally, an eavesdropper could learn that \emph{some} honest user is browsing the Web or using VoIP, but not \emph{which} honest user.
		Yet, this point is fairly orthogonal to the design of \prifi and can be addressed by adding padding and/or dummy traffic, at the cost of higher bandwidth usage, as proposed by substantial related work~\citep{wright2009traffic, mathewson2004practical,winter2013scramblesuit, wang2014effective, dyer2012peek}.
	\item External sender/receiver anonymity:
\prifi's anonymity set consists of the LAN users connected to a relay.\@
		Users outside the LAN are not anonymous.
		If both sender and receiver are part of a \prifi LAN, not necessarily the same, the protocol has sender and receiver anonymity.
	\item Intersection attacks, which correlate users' presence on the \prifi network with messages or other users,
are a practical threat to almost all ACNs~\citep{danezis2004statistical,wolinsky2013hang}. Although \prifi has no perfect solution to this problem, we discuss mitigation in Section~\ref{sec:intersections}, after presenting the system.
\squishend

\subsection{\prifi Solution Overview}

\prifi starts with a setup phase where clients authenticate themselves to the relay.
Clients and \servers then derive shared secrets.
Finally, clients are organized in a \emph{schedule} (a secret permutation) to decide when they communicate.

\para{Upstream Traffic.} 
The clients and the \servers run a DC-net protocol. 
Communication occurs in short \emph{time slots}. 
In each time slot, each client and \server sends a ciphertext to the relay.
The \emph{slot owner} can additionally embed some payload.
The relay waits for all ciphertexts, then computes the anonymized output.
This reveals one or more IP packet(s) without source address; the relay replaces it with its own IP address (as in a NAT) and forwards it to its destination.

Due to the construction of the DC-net, this protocol ensures provable anonymity. Ciphertexts are indistinguishable from each other to the adversary. During a slot, each client sends exactly the same number of bits. Finally, if the contribution from any honest client is missing, the output is undecipherable. Informally, this property achieves our goal G1 (Section~\ref{subsec:system-model-goals}) for upstream traffic.

\para{Precomputation of Ciphertexts.} The ciphertexts from the \servers are independent of the anonymous payloads. Hence, a key optimization is that \servers' ciphertexts are batch computed and sent in advance to the relay. The relay buffers and pre-combines the ciphertexts from multiple \servers, storing a single stream of pseudo-random bits to be later combined with clients' ciphertexts. This enables \prifi to have low latency despite the presence of high-latency links between the \servers and the relay.

\para{Latency-Critical Path.} Another important advantage of combining locally the ciphertexts is that clients' packets remain on their usual network path. The added latency is due mostly to the relay's need to wait for all clients. Similar systems that route clients' traffic between servers distributed around the Internet incur much higher latency.

\para{Downstream Traffic.} When receiving an answer to an anonymous message sent in some time slot, the relay encrypts it under the (anonymous) slot owner's public key, then broadcasts the ciphertext to all clients. As each client receives exactly the same message, this achieves our goal G1 (Section~\ref{subsec:system-model-goals}) for downstream traffic.

For broadcasting the downstream message, rather than performing $n$ unicast transmissions, the relay exploits the LAN topology and uses UDP broadcast, letting layer-2 network equipment (\eg switches) replicate the message if needed.
In WLANs, such a broadcast requires only one message, achieving receiver anonymity at no bandwidth or energy cost
in the absence of link-layer retransmissions.

	\section{Basic \prifi Protocol}
\label{sec:protocol-prifi}

\subsection{Preliminaries}

Let $\lambda$ be a standard security parameter, and let $\mathbb{G}$ be a cyclic finite group of prime order where the Decisional Diffie-Hellman (DDH) assumption~\cite{boneh1998decision} holds (\eg an elliptic curve). 

Let $(\texttt{KeyGen}, \mathcal{S}, \mathcal{V})$ be a signature scheme, with $\texttt{KeyGen}(\mathbb{G}, 1^\lambda)$ 
an algorithm that generates the private-public key pair $(p, P)$ used for signing.
We denote as $\sig{p}{m}$ the signature of the message $m$ with the key $p$.

Let $\texttt{KDF}:\mathbb{G}(1^\lambda) \to \bin^\lambda$ be a key derivation function that converts a group element into a bit string that can be used as a symmetric key.
Let $(\mathcal{E}, \mathcal{D})$ be a symmetric nonce-based encryption scheme~\citep{rogaway04nonce}.
We denote as $\enc{k}{m}$ the encryption of the message $m$ with the key $k$.
 
Let $H: \bin^* \to \bin^\lambda$ be a standard cryptographic hash function.
Let $\texttt{PRG}:\bin^\lambda\to \bin^*$ be a standard pseudo-random generator.
Let $F_1:\bin^{\lambda}\to\mathbb{G}$ be a public, invertible mapping function from binary strings to $\mathbb{G}$, and let $F_2:\bin^*\to\mathbb{G}$ be a hash function that maps bitstrings of arbitrary length to any point in $\mathbb{G}$ with uniform probability (\eg Elligator Squared \citep{tibouchi2014elligator}).
Finally, let $F_3:\bin^*\to\mathbb{N}$ be a public function that maps bitstrings to integers. 

\para{Identities.} Each party has a long-term key pair (denoted with the \emph{hat} symbol) generated with $\texttt{KeyGen}(\mathbb{G}, 1^\lambda)$:
\squishlist
	\item $(\hat{p}_i,\hat{P}_i)$ for client $C_i$, with $i\in \{1,\ldots,n\}$
	\item $(\hat{p}_j,\hat{P}_j)$ for \server $S_j$, with $j\in \{1,\ldots,m\}$
	\item $(\hat{p}_r,\hat{P}_r)$ for the relay
\squishend

Let $\vec{v}$ be the vector notation for $v$.
For each epoch, the group definition $G$ consists of all long-term public keys $G=(\vec{\hat{P}_i},\vec{\hat{P}_j},\hat{P}_r),~i\in\{1,\ldots,n\},~j\in\{1,\ldots,m\}$, and $G$ is known to all parties (\eg via a public-key infrastructure).
Finally, let $T=(\hat{P}_{1}, \hat{P}_{2}, \ldots)$ be a static roster of allowed clients known to the relay and the clients (\eg via a configuration file).

\subsection{Protocols}
\label{subsec:protocol-prifi-phases}

\prifi starts with the protocol \setup (Protocol~\ref{alg:setup}), followed by several runs of the protocol \anonymize (Protocol~\ref{alg:prifi}).

\subsubsection{Setup}

Each client authenticates itself to the relay using its long-term public key, and generates a fresh ephemeral key-pair.
Then, each client $C_i$ runs an authenticated Diffie-Hellman key exchange protocol with each \server $S_j$, using the fresh key pair to agree on a shared secret $r_{ij}$.
This secret is used later to compute the DC-net's ciphertexts.

Then, to produce a permutation $\pi$, the \servers shuffle the client's ephemeral public keys $\vec{P_i}$ by using a verifiable shuffle (\eg Neff's verifiable shuffle~\citep{neff2003verifiable}).
The public keys in $\pi$ correspond to the keys in $\vec{P_i}$, in a shuffled order, such that no one knows the full permutation.
Only a client holding the private key $p_i$ corresponding to an input in $\vec{P_i}$ can recognize the corresponding pseudonym key in $\pi$.

\begin{algorithm}
	\caption{\setup}
	\label{alg:setup}
	\small
	
	\begingroup
	\leftskip-1em
	\rightskip\leftskip
	
	{\bf Inputs:} $\lambda, \mathbb{G}, G, T$
	\vs{0.1cm}
	
	{\bf Outputs:} schedule $\pi$, shared secrets $r_{ij}$ between each client/\server pair $(C_i,~S_j)$
	\vs{0.25cm}

	{\bf 1. Client$\to$Relay Auth.} Each client $C_i$ generates a fresh key pair $(p_i, P_i) \gets \texttt{KeyGen}(\mathbb{G}, 1^\lambda)$ and sends $P_i,\sig{\hat{p}_i}{P_i}$ to the relay.
	The relay checks the signature and that $\hat{P}_i\in T$, and it replies with $\sig{\hat{p}_r}{P_i}$.
	\vs{0.1cm}
	
	{\bf 2. Client$\to$\Server Auth.} Each client $C_i$ sends $P_i,\sig{\hat{p}_i}{P_i},\sig{\hat{p}_r}{P_i}$ to all \servers.
	\vs{0.1cm}
	
	{\bf 3. Shared Secrets.} Each \server $S_j$ derives $n$ secrets $r_{ij}=\texttt{KDF}(\hat{p}_j\cdot P_i)$, one for each client with a valid signature $\sig{\hat{p}_r}{P_i}$ from the relay.
	Similarly, each client $C_i$ derives $m$ secrets $r_{ij}=\texttt{KDF}(p_i\cdot \hat{P}_j)$.
	\vs{0.1cm}
	
	{\bf 4. Verifiable Shuffle.} Clients participate in a verifiable shuffle protocol~\citep{neff2003verifiable} run by the \servers, with the ephemeral keys $\vec{P_i}$ as input.
	The public output $\pi$ consists of $n$ pseudonym keys in permuted order, such that no one knows which client corresponds to which key, except the owner of the corresponding private key.
	More formally, we write $\pi=(\tilde{P}_{\alpha{(1)}}, \ldots, \tilde{P}_{\alpha{(n)}})$, where $\tilde{P}_{\alpha{(i)}} = c\cdot P_i$ for a permutation $\alpha$ and some constant $c$.
	At the end of this step, clients receive $\pi$, along with a transcript signed by all \servers.
	\vs{0.1cm}
	
	{\bf Safety Checks.} In step 4, the honest \server checks that each input $P_i$ corresponds to a client with a valid $\sig{\hat{p}_r}{P_i}$, or it aborts.
	
	{ At the end of \setup, honest clients check that (1) the verifiable shuffle completed correctly, (2) $\pi$ is signed by every \server in $G$, (3) there are at least $K=2$ clients in $T$ in the input, and (4) its own shuffled pseudonym is included in the permutation. If any test fails, it aborts. }
	\vs{0.2cm}
	
	{ Finally, the relay creates $n$ empty dictionaries $b_k$, indexed by $k=\alpha(i)$, to keep track of IP sockets later used for packet forwarding. }
	
	\endgroup
\end{algorithm}

\vs{0.1cm}
\noindent \setup has the following properties (proved in~\ref{subsec:proofs-setup}):

\para{Property~\ref{thm1}.} A shared secret $r_{ij}$ between an honest client $C_i$ and an honest \server $S_j$ is known only to $C_i$ and $S_j$.

\para{Property~\ref{thm2}.} Let $C_0$ and $C_1$ be two honest clients who ran \setup without aborting, and $\alpha(0), \alpha(1)$ the position of their respective shuffled key in $\pi$.
Then, the adversary $\ad$ has negligible advantage in guessing $b\in[0,1]$ such that the mapping client $\to$ position $(b\to \alpha(0), (1-b)\to \alpha(1))$ is in $\pi$. 

\para{Remark.} The setup protocol (client/server secret sharing with a verifiable shuffle of client pseudonyms) is similar to that used in closely-related work~\citep{corrigan2010dissent,wolinsky2012dissent}.

\subsubsection{Anonymize}

After \setup, all nodes continuously run \anonymize.
In each time slot, clients and \servers participate in a DC-net protocol.
All \servers compute one  $\ell$-bit pseudo-random message from the $\texttt{PRG}$s seeded with the shared secrets, and send it to the relay.
All clients perform likewise, except for the client owning the time slot, who additionally includes its upstream message(s) $m_i$ in the computation.
In practice, $m_i$ is one or more IP packet(s) without source address, up to a total length $\ell$.
If the slot owner has nothing to transmit, it sets $m_i=0^\ell$.

Once the relay receives the $n+m$ ciphertexts from all clients and \servers, it XORs them together to obtain $m_k$.
If the protocol is executed correctly, $m_k$ is equal to $m_i$, as the values of  $\texttt{PRG}(r_{ij}),i\in\{1\ldots,n\},j\in\{1\ldots,m\}$ cancel out.
If $m_k$ is a full IP packet, the relay replaces the null source IP in the header by its own (just like in a NAT) and forwards it to its destination.
If it is a partial packet, the relay buffers it and completes it during the next schedule.

Then, the relay broadcasts one downstream message $\vec{d}$ to all clients, each $d\in\vec{d}$ being encrypted with a public key $\tilde{P}_k \in \pi$ corresponding to an anonymous client. We emphasize that the relay does not know for which client it encrypts. Additionally, $\vec{d}$ is of arbitrary length $\ell'$, possibly much larger than $\ell$, easily accommodating downstream-intensive scenarios.
Finally, we emphasize that $\vec{d}$ can contain data for multiple users (from previous rounds). If the relay has nothing to transmit, it sends a single $0$ bit to indicate the end of the round.

\begin{algorithm}
	\caption{\anonymize}
	\label{alg:prifi}
	\small
	
	\begingroup
	\leftskip-1em
	\rightskip\leftskip
	
	{\bf Inputs:} $\vec{r_{ij}},\pi$, up/down-stream message sizes $\ell,\ell'$
	\vs{0.1cm}
	
	{\bf Outputs:} per round $k$: anonymous message $m_k$, downstream traffic $\vec{d}$.
	\vs{0.25cm}
	
	For round $k\in\{1,\ldots,n\}$:
	\begin{itemize}[leftmargin=0pt, rightmargin=2.5em]
		\item Each client $C_i$ sends to the relay 
		$c_i \gets \texttt{DCNet-Gen}(\vec{r_{ij}}, x_i)$ with
		
		$x_i=\begin{cases}
		m_i,& \text{if } \alpha(i)=k,{\color{gray}\text{\hspace{1cm}//}C_i\text{ is the sender}}\\
		\vec{0},              & \text{otherwise}.
		\end{cases}$
		
		\item Each \server $S_j$ sends to the relay\\
		$s_j \gets \texttt{DCNet-Gen}(\vec{r_{ij}},\vec{0})$.
		
		\item The relay computes $m_k\gets \texttt{DCNet-Reveal}(\vec{c_i}, \vec{s_j})$, with $m_k \in {\{0,1\}}^\ell$ an IP packet.
		
		\item The relay handles $m_k$ as follows:
		
		\begin{itemize}[leftmargin=*, rightmargin=0pt]
			\item If $m_k$ is not part of an active socket in $b_k$, the relay creates and stores the socket.
			\item it puts its own IP address in $m_k$, then sends it in the appropriate socket in $b_k$.
		\end{itemize}
		
		\item The relay computes\\
		$\vec{d} \gets \texttt{Downstream}(\vec{b})$ and sends $\vec{d}$ to each client.
	\end{itemize}\vs{-0.2cm}	
	
	\SetAlgoNoLine
	\DontPrintSemicolon
	\SetKwProg{Fn}{Function}{:}{}
	
	\SetKwFunction{FMain}{DCNet-Gen}
	\Fn{\FMain{$\vec{r_{ij}}, x_i$}}{
		\KwRet $\bigoplus_{r \in \vec{r_{ij}}} \texttt{PRG}(r)~\oplus~x_i$
	}
	\vs{0.1cm}
	
	\SetKwFunction{FMain}{DCNet-Reveal}
	\Fn{\FMain{$\vec{c_i}, \vec{s_j}$}}{
		\KwRet $\bigoplus_{i}c_i~\oplus~\bigoplus_{j}s_j$
	}
	\vs{0.1cm}
	
	\SetKwFunction{FMain}{Downstream}
	\Fn{\FMain{$\vec{b}$}}{
		$\vec{d} \gets $ array(); \;
		\For{$k\in \{1,\ldots,n\}$}{
			\For{socket $\in b_k$ \emph{containing downstream bytes} $d$}{
                add $\enc{\tilde{P}_k}{d}$ to $\vec{d}$
            }
		}
	}	
	\endgroup
\end{algorithm}\vs{-0.5cm}

\vs{0.1cm}
\noindent \anonymize has the following property (proved in~\ref{subsec:proofs-anonymize}):

\para{Property~\ref{thm3} [Goal G1].} After a run of \anonymize, let $C_{i_1}$ and $C_{i_2}$ be two honest clients, $k_1=\alpha(i_1), k_2=\alpha(i_2)$ the time slots in which they communicated, and $m_{k_1},m_{k_2}$ the anonymous upstream messages for those slots.
Then, $\mathcal{A}$ has negligible advantage in guessing $b\in[1,2]$ such that $m_{k_b}$ is the message sent by $i_1$.

\subsection{Practical Considerations}
\label{subsec:protocol-prifi-practical-considerations}

\para{End-to-End Confidentiality.} A malicious relay can see the upstream message plaintexts.
This is also the case for a VPN server or Tor exit relay; clients should use standard end-to-end encryption (\eg TLS) on top of \prifi. 

\para{Churn.} In the case of churn, \eg if any client or \server joins or disconnects, the relay broadcasts a \setup request that signals the start of a new epoch.
Upon reception, each node aborts and re-runs \texttt{Setup}.
Churn negatively affects performance; we evaluate its effect in Section~\ref{subsec:eval-client-churn}.

\para{Bandwidth Usage.} To reduce idle bandwidth usage, the relay periodically sends a ``load request'' in which clients can anonymously \emph{open} or \emph{close} their slots.
The relay skips a closed slot, hence saving time and bandwidth. If all slots of a schedule are closed, the relay sleeps for a predetermined interval, further saving bandwidth at the cost of higher initial latency when resuming communications.
The concrete parameters of this improvement (\eg frequency of the load requests, sleep time) are not fully explored in this work; they exhibit a typical latency-bandwidth usage trade-off.

We emphasize that load tuning does not reduce the anonymity set size; all clients still transmit ciphertexts exactly at the same time.
Load tuning makes global communication volumes and packet timings more visible to an external eavesdropper, but our threat model considers a stronger, local eavesdropper (the malicious relay) who has access to this information anyway.

Finally, although this has not been investigated in this work, both up/down-stream sizes $\ell$ and $\ell'$ can be dynamically tuned without interrupting the communications. This allows the relay to better match the sending/receiving rates of the clients and further reduce idle bandwidth usage.

\para{Synchronicity.} The protocol uses message reception events, rather than clocks, to keep the participants in lock-step.

\subsection{Limitations of this Protocol}

\para{Accountability.} No mechanism enforces dishonest parties to correctly follow the protocol; malicious parties can anonymously disrupt the communications.
This is a well-known DC-net issue \citep{chaum1988dining,corrigan2010dissent,wolinsky2012dissent}, addressed in Section~\ref{sec:protocol-disruption}.

\para{Downstream Anonymity.} This notion refers to the clients being indistinguishable when receiving downstream messages, which is trivially the case if the relay truthfully sends the same downstream data to all clients.
This property is not enforced above,
but is addressed later in Section~\ref{sec:protocol-equivocation}.

	\section{Disruption Protection}
\label{sec:protocol-disruption}

In the basic protocol above, a malicious active insider can modify or jam upstream communications by transmitting arbitrary incorrect bits
instead of the ciphertext defined by the protocol.
This is particularly problematic because the attacker is provably anonymous and untraceable.

In the related work, these attacks can be detected retroactively using ``trap bit'' protocols~\citep{wolinsky2012dissent} that detect a disruptor with a certain probability but reduce the throughput linearly with respect to the number of trap bits. Unfortunately, the probability of detection must be high enough to detect a single bit-flip, which can effectively corrupt a message. Another technique is to rely on commitments before every DC-net message~\citep{corrigan2010dissent}, which adds latency.

Some DC-nets use group arithmetic instead of binary strings, which enables proving (proactively or retroactively) that computations are correct~\citep{corrigan2013verdict,golle2004dining}. These designs do not fit low-latency requirements, unfortunately, as their computation time is significantly higher: tens of milliseconds per message for the computation alone, whereas XOR-based DC-nets take microseconds. A brief analysis of this computational cost is provided in Verdict~\citep{corrigan2013verdict} (p.$12$, Figure~$6$).

\prifi uses a retroactive, hash-based ``blame'' mechanism on top of a ``classic'' XOR-based DC-net, which (1) keeps the typical operation (in the absence of jamming) as fast as possible, and reduces the bandwidth lost due to the protection, and (2) excludes a disruptor with high probability ($1/2$ per flipped bit). Exploiting the LAN topology, our exclusion takes seconds, which is orders of magnitude faster than the related work~\citep{wolinsky2012dissent} (see Figure~\ref{fig:disruption-blame}).

\subsection{Protocol}

We modify the previous \anonymize protocol (Protocol~\ref{alg:prifi}) to protect against disruption from malicious insiders.
In short, we add a hash-based detection of corruption and a blame mechanism to exclude a disruptor.

\para{Summary.} The relay sends the hash of the upstream message on the downstream traffic.
If the anonymous sender detects an incorrect hash, it requests a copy of its own disrupted message by setting a flag $b_\text{echo\_last}$ to $1$.

When receiving a disrupted copy $m_k'$ of a previously-sent message $m_k$, the client searches for a bit position $l$ such that ${(m_k)}_l=0$ and ${(m_k')}_l=1$.
If such $l$ exists, then the client requests to de-anonymize the $l^\text{th}$ bit of his own slot $k$, by sending $\text{NIZKPoK}_{k,l}(\tilde{p}_k: \tilde{p}_k=\log~\tilde{P}_k)$ in its next upstream message, a non-interactive proof of knowledge of the key $\tilde{p}_{(k~\text{mod}~n)}$ corresponding to slot $k$ in $\pi$~\citep{blum1988non}.
The proof is bound to the public values $l$ and $k$. For simplicity, we write $\text{PoK}_{k,l}(\tilde{p}_k: \tilde{p}_k=\log~\tilde{P}_k)$ hereafter, thus ignoring (1) the mod $n$ and (2) the acronym for ``Non-Interactive, Zero-Knowledge''.

If no such $l$ exists, the message was disrupted but the disruptor cannot be traced without simultaneously de-anonymizing a client (see ``Remarks'' below for more details).
In this case, nothing happens.


\vs{0.1cm}
\noindent The \anonymize and \blame protocols are described in Protocols~\ref{alg:prifi2} and~\ref{alg:disruption-blame}, respectively, and have the following properties (proved in Appendix~\ref{subsec:proofs-disruption}):

\begin{algorithm}
	\caption{\anonymize}
	\label{alg:prifi2}
	\small
	
	\begingroup
	\leftskip-1em
	\rightskip\leftskip
	
	\setlist[enumerate]{rightmargin=3.3em}
	
	(The differences with Protocol~\ref{alg:prifi} are highlighted in \hl blue\bk.)
	\vs{0.1cm}
	
	{\bf Inputs:} $\vec{r_{ij}},\pi,\ell,\ell'$
	\vs{0.1cm}
	
	{\bf Outputs:} per round $k$: $m_k$, $\vec{d}$.
	\vs{0.25cm}
	
	For round $k\in\{1,\ldots,n\}$:
	\begin{itemize}[leftmargin=0pt]
		\item Each client $C_i$ sends to the relay \\
		$c_i \gets \texttt{DCNet-Gen}(\vec{r_{ij}}, x_i)$ with
		
		$x_i=\begin{cases}
		\vec{0},              & \text{if } \alpha(i)\neq k,\\
		\hl \text{PoK}_{{k'},l}(\tilde{p}_{k'}: \tilde{p}_{k'}=\log~\tilde{P}_{k'}),              & \hl \text{if slot}~k'~\text{was disrupted},\\
		m_i\hl||b_\text{echo\_last}\color{black},& \text{otherwise}.
		\end{cases}$
		
		\item Each server $S_j$ sends to the relay $s_j \gets \texttt{DCNet-Gen}(\vec{r_{ij}},\vec{0})$.
		
		\item The relay computes $m_k\gets \texttt{DCNet-Reveal}(\vec{c_i}, \vec{s_j})$
		
		\item The relay handles $m_k$ as follows:
		
		\begin{itemize}[leftmargin=*]
			\item \hl if $m_k$ is a Blame message, it starts\\
			the $\blame(\text{PoK}_{{k'},l}(\tilde{p}_{k'}: \tilde{p}_{k'}=\log~\tilde{P}_{k'}),\vec{c_i},\vec{s_j})$ protocol, \color{black}
			\item else, it sends $m_k$ in the appropriate socket in $b_k$.
		\end{itemize}
		
		\item The relay computes and sends to each client\\
		$\vec{d} \gets \hl\text{Downstream2}(\vec{b},m_k,k,b_\text{echo\_last})\color{black}$
	\end{itemize}\vs{-0.2cm}
	
	\SetAlgoNoLine
	\DontPrintSemicolon
	\SetKwProg{Fn}{Function}{:}{}
		
	\SetKwFunction{FMain}{Downstream2}
	\Fn{\FMain{$\vec{b},m_k,k,b_\text{echo\_last}$}}{
		$\vec{d} \gets $ array();\;
		\For{$k'\in \{1,\ldots,n\}$}{
			\vs{-3mm}\begin{enumerate}[leftmargin=*]
				\item \hl if $k'=k$:~~(for the anonymous sender)
				\begin{enumerate}[leftmargin=*]
					\item add $H(m_k)$ to $\vec{d}$,
					\item if $b_\text{echo\_last} = 1$: add $\enc{\tilde{P}_{k}}{m_{k-n}}$ to $\vec{d}$.
				\end{enumerate}\color{black} 
				\item for each socket $\in b_{k'}$ containing downstream\\
				 bytes $d$, add $\enc{\tilde{P}_{k'}}{d}$ to $\vec{d}$.
			\end{enumerate}
		\vs{-5mm}}
	}
	\endgroup
\end{algorithm}

\begin{algorithm}
	\caption{\blame}
	\label{alg:disruption-blame}
	\setlist[enumerate]{rightmargin=2.5em}
	
	\begingroup
	\leftskip-1em
	\rightskip\leftskip
	\small 
	
	{\bf Inputs:} $\text{PoK}_{k,l}(\tilde{p}_k: \tilde{p}_k=\log~\tilde{P}_k),\vec{c_i},\vec{s_j}$
	\vs{0.1cm}

	\begin{enumerate}[leftmargin=0pt]
		\item The relay broadcasts $\text{PoK}_{k,l}(\tilde{p}_k: \tilde{p}_k=\log~\tilde{P}_k)$\\ to every client/\server.
		\item Each client/\server checks the PoK, and reveals the $l^\text{th}$ bit of the values $\texttt{PRG}(r_{ij})$ for slot $k$, $\forall i\in\{1,\ldots,n\},j\in\{1,\ldots,m\}$ by sending a non-anonymous, signed message $\vec{{\texttt{PRG}(r_{ij})}_l},\sig{\hat{p}_i}{\vec{{\texttt{PRG}(r_{ij})}_l}}$ to the relay. A non-complying entity is identified as the disruptor.
		\item For each client, the relay checks the signature, and that $\bigoplus_j {\texttt{PRG}(r_{ij})}_l$ indeed equals ${(c_i)}_l$ sent in slot $k$; if a mismatch is detected, this client is identified as the disruptor. The relay performs the same verification for each \server.
		\item For each pair of client-\server $(C_i,S_j)$, the relay compares ${\texttt{PRG}(r_{ij})}_l$ from the client and ${\texttt{PRG}(r_{ij})}_l$ from the \server: they should be equal. If there is a mismatch, at least one of them is lying. The relay continues by forwarding the signed message $\vec{{\texttt{PRG}(r_{ij})}_l}$ from $C_i$ to $S_j$ and vice-versa.
		\item $C_i$ checks that $\vec{{\texttt{PRG}(r_{ij})}_l}$ is signed by $S_j$, and that the bit ${\texttt{PRG}(r_{ij})}_l$ is in contradiction with its own bit ${\texttt{PRG}(r_{ij})}_l$. Then, he answers with $r_{ij}$, the secret shared with $S_j$, along with a proof of correctness for computing $r_{ij}$. $S_j$ proceeds similarly. A non-complying entity is identified as the disruptor.
		\item The relay checks the proofs of correctness, then uses $r_{ij}$\\ to recompute the correct value for $\texttt{PRG}(r_{ij})$ for slot $k$,\\ identifying the disruptor.
	\end{enumerate}
	\vs{-0.3cm}
	\begin{adjustwidth}{-0.3cm}{0.8cm}
		Once the disruptor is identified, the relay excludes it from the group $G$ and the roster $T$, and then broadcasts all inputs and messages exchanged in \blame to all clients for accountability.
	\end{adjustwidth}
	\endgroup
\end{algorithm}

\para{Properties~\ref{thm4}+\ref{thm5} [Goal G1].} The anonymity of any honest client is unaffected by the information made public in \blame (Protocol~\ref{alg:disruption-blame}): $\vec{{\texttt{PRG}(r_{ij})}_l}$  in step $2$, or $r_{ij}$ in step $5$.

\para{Property~\ref{thm7} [Goal G2].} Let $C_i$ be the owner of a slot $k$, and let $C_d,~d\neq i$, be another client (or \server).
If $C_d$ sends an arbitrary value $q$ instead of the value $c_i$ (or $s_j$) as specified in the protocol, then $C_d$ is identified as the disruptor and is excluded from subsequent communications.  

\para{Property~\ref{thm6}.} An honest entity is never identified as a disruptor. 

\para{Limitations.} The detection relies on the capacity of the jammed client to transmit $b_\text{echo\_last}$ and the PoK\@; the adversary also can jam these values. In practice, $l$ is fairly large (\eg $5$~kB), and the client can use redundancy coding over his message to make the task harder for the adversary. When this probabilistic solution is insufficient, users can use a verifiable DC-net~\citep{corrigan2013verdict} to transmit without the risk of jamming; in practice, this verifiable DC-net would run in background with very low bandwidth and high latency, just enough to transmit the proof-of-knowledge.

\para{Remarks.} In step $3$ of \blame (Protocol~\ref{alg:disruption-blame}), we observe why the client starting the blame checks that ${(m_k)}_l=0$: otherwise, revealing $\vec{{\texttt{PRG}(r_{ij})}_l}$ over a non-anonymous channel would flag this client as the sender, as $\bigoplus_j {\texttt{PRG}(r_{ij})}_l\neq {(m_k)}_l$.

In step $5$ of \blame (Protocol~\ref{alg:disruption-blame}), we remark that at least one mismatching pair exists, otherwise the slot would not have been disrupted. If multiple disruptors exist,
\blame excludes one disruptor per disruption event.

	\section{Equivocation Protection} 
\label{sec:protocol-equivocation}

In both versions of \anonymize above (Protocols~\ref{alg:prifi} and~\ref{alg:prifi2}), the relay broadcasts the downstream traffic $\vec{d}$ to all clients to ensure receiver anonymity.
However, no mechanism enforces truthful broadcast, so a malicious relay can perform \emph{equivocation attacks}: \ie send different messages to each client, hoping that their subsequent behaviors will reveal which client actually decrypted the message.
This attack can be seen as a ``poisoning'' of downstream traffic.


\para{Equivocation Example.} Clients $C_1$ and $C_2$ are both honest.
On the first round, the relay decodes an anonymous DNS request.
Instead of broadcasting the same DNS answer to $C_1$ and $C_2$, the relay sends two different answers containing $\text{IP}_1$ and $\text{IP}_2$, respectively.
Later, the relay decodes an anonymous IP packet with $\text{IP}_2$ as destination.
It can guess that $C_2$ made the request, as $C_1$ has never received $\text{IP}_2$.


In practice, a credible scenario is a router infected with malware or compromised by the adversary and spying on honest users of a corporate network, possibly colluding with the endpoints contacted by the clients.

We note that previous DC-net systems do not mention this issue~\citep{corrigan2010dissent,wolinsky2012dissent,corrigan2013verdict}. The attack is possible because a malicious party relays the information, which can happen in both Dissent~\cite{wolinsky2012dissent} and Verdict~\cite{corrigan2013verdict}.
If the traffic is unencrypted, the attack is trivial. The use of higher-level encryption (\eg TLS) can offer a mitigation, \emph{if} we further assume that the remote endpoint does not collude with the malicious relay. 
In practice, having two particular entities under the control of the adversary (the relay and some interesting external service, \eg WikiLeaks)  does not seem impossible, however.

On the contrary, we note that due to their different design, mix-nets and onion-router networks are typically not affected by this attack.


\para{\prifi Solution.} Intuitively, to thwart an equivocation attack, clients need to agree on what they have received before transmitting their next message.
In \prifi, this is achieved without adding extra latency and without synchronization between clients.
Clients encrypt their messages before anonymizing them;
the encryption key depends on the history of downstream messages and also on the shared secrets with the \servers. The relay is thus unable to recover a plaintext if not all clients share the same history.

\subsection{Protocol}

We modify the previous \anonymize and \blame protocols (Protocol~\ref{alg:prifi3} and~\ref{alg:equiv-blame}). These are the final variants used in our implementation (Section~\ref{sec:evaluation}).

\para{\anonymize.} Each client $C_i$ keeps a personal history $h_i$ of downstream communications.
Upon receiving a downstream message $\vec{d}$, each client updates its history.

Each upstream message is then symmetrically-encrypted with a fresh key $\gamma$.
This value $\gamma$ is sent to the relay, blinded by a function of the downstream history $h_i$. Only if all honest clients agree on the value $h_i$, the relay can unblind $\gamma$ and decrypt the message.

\para{\blame.} The previous \blame protocol finds a disruptor when values $c_i$ or $s_j$ are not computed correctly; we add a way to validate the new values $\kappa_i$ and $\sigma_j$. As they are elements of $\mathbb{G}$, we apply a standard zero-knowledge proof~\citep{blum1988non}.
More precisely, we use an Or/And type of NIZKPoK which allows the clients to prove either (1) their ownership of the slot or (2) that $\kappa_i$ is computed correctly.

\vs{0.1cm}
\noindent The \anonymize and \blame protocols have the following properties (proved in Appendix~\ref{subsec:proofs-equivocation}):

\begin{algorithm}
	\caption{\anonymize~{\small(final version)}}
	\label{alg:prifi3}
	\small
	
	\begingroup
	\leftskip-1em
	\rightskip\leftskip
	
	\setlist[enumerate]{rightmargin=3.5em}
	
	(The differences with Protocol~\ref{alg:prifi2} are highlighted in \hl blue\bk.)
	\vs{0.1cm}
	
	{\bf Inputs:} $\vec{r_{ij}},\pi,\ell,\ell'$
	\vs{0.1cm}
	
	{\bf Outputs:} per round $k$: $m_k$, $\vec{d}$.
	\vs{0.25cm}
	
	For round $k\in\{1,\ldots,n\}$:
	\begin{itemize}[leftmargin=0pt,rightmargin=2.5em]
		\item Each client $C_i$ sends to the relay \\
		$\hl c_i, \kappa_i \gets \texttt{DCNet-Gen-Client}(\vec{r_{ij}}, m_i, h_i)$ 
		
		\item Each \server $S_j$ sends to the relay \\
		$\hl s_j, \sigma_j \gets \texttt{DCNet-Gen-\Server}(\vec{r_{ij}})$

		\item The relay computes \\ $m_k\gets \hl\texttt{DCNet-Reveal2}(\vec{c_i},\vec{s_j},\vec{\kappa_i},\vec{\sigma_j})\color{black}$ with $m_k \in \bin^l$ or $\hl\bot\bk$.
		
		\item The relay handles $m_k$ as follows:
		
		\begin{itemize}[leftmargin=*]
			\item if $m_k$ is a Blame message, it starts the $\blame$ protocol,
			\item else, it sends $m_k$ in the appropriate socket in $b_k$.
		\end{itemize}
		
		\item The relay outputs $\vec{d} \gets \texttt{Downstream2}(\vec{b},H(m_k),k,b_\text{echo\_last})$
		\item \hl The relay updates $h_r \gets H(h_r || \vec{d})$, and sends $\vec{d},\sig{\hat{p}_r}{h_r}$\\ to each client.
		\item \hl When receiving $\vec{d}$, each client checks the signature $\sig{\hat{p}_r}{h_r}$ or aborts. Then, each client $C_i$ updates $h_i \gets H(h_i || \vec{d})$. \bk
	\end{itemize}	

	\vs{-0.3cm}
	\SetAlgoNoLine
	\DontPrintSemicolon
	\SetKwProg{Fn}{Function}{:}{}
	
	\SetKwFunction{FMain}{DCNet-Gen-Client}
	\Fn{\FMain{$\vec{r_{ij}}, m_i, h_i$}}{\vs{-10pt}
		\begin{enumerate}[leftmargin=0pt,rightmargin=2.5em]
		\item compute $x_i$ as follows:
		\begin{itemize}[leftmargin=1.5em,rightmargin=2.5em]
			\item if $\alpha(i)=k$:
			\begin{itemize}[leftmargin=3em,rightmargin=2.5em]
				\item if slot $k'$ was disrupted, $x_i\gets \text{PoK}_{k,l}(\tilde{p}_k: \tilde{p}_k=\log~\tilde{P}_k)$
				\item \hl else, pick a random symmetric key $\gamma \overset{\$}{\gets} \bin^{\lambda}$, compute $m_i'=\enc{\gamma}{m_i}$, and set $x_i\gets m_i'||b_\text{echo\_last}$
			\end{itemize}
			\item else, $x_i\gets \vec{0}$
		\end{itemize}
		\item compute $c_i \gets \bigoplus_{i} \texttt{PRG}(r_{ij})~\oplus~x_i$
		\item \hl compute $\kappa_i$ as follows:
		\begin{itemize}[leftmargin=1.5em,rightmargin=2.5em]
			\item if $\alpha(i)=k$: 
			
			$\kappa_i \gets F_1(\gamma) + F_2(h_i)\cdot\sum_{j=1}^{m}F_3(H(\texttt{PRG}(r_{ij})))$
			\item else,
			
			$\kappa_i \gets \color{white}F_1(\gamma) + \hl F_2(h_i)\cdot\sum_{j=1}^{m}F_3(H(\texttt{PRG}(r_{ij})))$
		\end{itemize}
	\end{enumerate}\vs{-8pt}
	\KwRet $c_i, \hl \kappa_i$
	}
	\vs{0.1cm}
	
	\SetKwFunction{FMain}{DCNet-Gen-\Server}
	\Fn{\FMain{$\vec{r_{ij}}, x_i$}}{
		$s_j \gets \bigoplus_{i} \texttt{PRG}(r_{ij})$ \;
		$\hl \sigma_j \gets -\sum_{i=1}^{n}F_3(H(\texttt{PRG}(r_{ij})))$ \;
		\KwRet $s_j, \hl \sigma_j$
		
	}
	\vs{0.1cm}
	
	\SetKwFunction{FMain}{DCNet-Reveal2}
	\Fn{\FMain{$\vec{c_i}, \vec{s_j}$}}{
		$m_k' \gets \bigoplus_{i}c_i~\oplus~\bigoplus_{j}s_j$ \;
		\hl $F_1(\gamma) \gets F_2(h_r)\cdot\sum_{j=1}^{m}\sigma_j~+~\sum_{i=1}^{n}\kappa_i$ \;
		$m_k \gets \dec{\gamma}{m_k'}$ \; \bk
		\KwRet $m_k$
	}
	\vs{0.1cm}

	\endgroup
\end{algorithm}

\para{Properties~\ref{thm8}+\ref{thm9}+\ref{thm10}+\ref{thm11} [Goal G1].} The anonymity of any honest client is unaffected by the extra information revealed: $\kappa_i$ in step $3$ of \texttt{DCNet-Gen-Client}, $\sigma_j$ in step $2$ of \texttt{DCNet-Gen-\Server}, $\vec{Q_i}, \text{PoK}$ in step $7$ of \blame, or $r_{ij}$ in step $9$ of \blame.

\para{Property~\ref{thm12} [Goal G1].} If $\exists i,j$ two honest clients who received $\vec{d}_i \neq \vec{d}_j$ on round $k$, then neither the relay nor $\ad$ can decrypt $m_k$ for any subsequent round $k'>k$.

\para{Property~\ref{thm13} [Goal G2].} If a client $C_i$ sends an arbitrary value $\kappa_i'$ instead of the value $\kappa_i$ as specified in the protocol, then $C_i$ is identified as the disruptor and is excluded from subsequent communications.

\para{Properties~\ref{thm14}+\ref{thm15}.} Honest parties are not blamed for equivocation attacks or for disruption attacks.

\begin{algorithm}
	\caption{\blame~{\small(final version)}}
	\label{alg:equiv-blame}
	\setlist[enumerate]{rightmargin=0em}
	\begingroup
	\leftskip-1em
	\rightskip\leftskip
	\small 
	
	(The differences with Protocol~\ref{alg:disruption-blame} are highlighted in \hl blue\bk.)
	\vs{0.1cm}
		
	{\bf Inputs:} $\text{epoch ID}~e,~\text{PoK}_{k,l}(\tilde{p}_k: \tilde{p}_k=\log~\tilde{P}_k),\vec{c_i},\vec{s_j}$
	\vs{0.1cm}
	
	\begin{enumerate}[leftmargin=0pt,rightmargin=25pt]
		\item The relay broadcasts $\text{PoK}_{k,l}(\tilde{p}_k: \tilde{p}_k=\log~\tilde{P}_k)$\\ to every client/\server,
		\item Each client/\server checks the PoK and sends to the relay $\vec{{\texttt{PRG}(r_{ij})}_l},\sig{\hat{p}_i}{e,\vec{{\texttt{PRG}(r_{ij})}_l}}$ for slot $k$, $\forall i\in\{1,\ldots,n\},j\in\{1,\ldots,m\}$.
		\item For each client/\server, the relay checks the signature, and that $\bigoplus_j {\texttt{PRG}(r_{ij})}_l$ indeed equals ${(c_i)}_l$ sent in slot $k$.
		\item For each pair of client-\server $(C_i,S_j)$, the relay compares ${\texttt{PRG}(r_{ij})}_l$ from the client and ${\texttt{PRG}(r_{ij})}_l$ from the \server. If there is a mismatch, the relay forwards the signed message $\vec{{\texttt{PRG}(r_{ij})}_l}$ from $C_i$ to $S_j$ and vice-versa.
		\item $C_i$ checks that the signature and that the bit ${\texttt{PRG}(r_{ij})}_l$ mismatches with its own bit ${\texttt{PRG}(r_{ij})}_l$.
			Then, it answers with $r_{ij}$ along with a proof of correctness.
			$S_j$ proceeds similarly.
		\item The relay checks the proofs, then uses $r_{ij}$ to recompute the\\correct value for $\texttt{PRG}(r_{ij})$ for slot $k$.
			\hl If no disruptor is\\
			identified, the relay continues.
		\item Each client $C_i$ computes $m$ values $Q_i=P\cdot F_3(H(\texttt{PRG}(r_{ij})))$, $\forall j\in\{1,\ldots,m\}$, where $P$ is the base point of $\mathbb{G}$.
		Then, each client computes a PoK as follows:
        \vs{-0.2cm}
		\begin{alignat*}{3}
		\text{PoK}\big\{\tilde{p}_k,q,q_1\ldots,q_m:\tilde{p}_k=\log~\tilde{P}_k \vee \{&q &&= \log~\kappa_i~&&\wedge\\
		&q_1 &&:= \log~Q_1~&&\wedge\\[-0.6em]
		&\vdots&& &&\\[-0.6em]
		&q_m &&:= \log~Q_m~&&\wedge\\
		&q &&:= q_1 + \cdots + q_m&&~\}\big\}
		\end{alignat*}
		
        \vs{-0.2cm}Each client sends a message $\vec{Q_i},\text{PoK},\sig{\hat{p}_i}{e,\vec{Q_i},\text{PoK}}$\\ to $R$.
		A non-complying entity is identified as\\ the disruptor.
		Each server performs similarly, minus the first clause of the PoK which is never true.
		
		\item The relay checks all signatures and PoKs, and potentially identifies a disruptor. If not, for each pair of client-\server $(C_i,S_j)$, the relay compares the two values $Q_i,Q_j$ --- they should be equal. 
		If there is a mismatch, at least one of them is lying.
		
		The relay continues by sending the signed message $\vec{Q_i},\text{PoK},\sig{\hat{p}_i}{e,\vec{Q_i},\text{PoK}}$ from $C_i$ to $S_j$ and vice-versa.
		\item $C_i$ checks the signature and the PoK, and that a value $Q_i$ is in contradiction with some value of its own.
		Then, it answers with $r_{ij}$, the secret shared with $S_j$, along with a proof of correctness for computing $r_{ij}$.
		$S_j$ proceeds similarly.
		A non-complying entity is identified as the disruptor.
		\item The relay checks the proofs of correctness, then uses $r_{ij}$ to recompute the correct values for $\texttt{PRG}(r_{ij})$ and $H(\texttt{PRG}(r_{ij}))$ for slot $k$, identifying the disruptor.
	\end{enumerate}
	\vs{-0.3cm}
	\begin{adjustwidth}{-0.3cm}{0.8cm}
	Once the disruptor is identified, the relay excludes it from the group $G$ and the roster $T$, and then broadcasts all inputs and messages exchanged in \blame to all clients for accountability.
	\end{adjustwidth}
	\endgroup
\end{algorithm}

\subsection{Practical Considerations}

\para{Packet Losses.} We note that this protection is decoupled from link-layer retransmissions; if one client fails to receive a packet, it will ask the relay again after a timeout, delaying all clients for this specific round (which is unavoidable for traffic-analysis resistance) but not invalidating the whole epoch with a wrong $h_i$ value.

\ifextended{}
\para{Remark.} Due to the ``Or'' format of the PoK, we note that a malicious client can jam their own slot, then send arbitrary $Q_i$ values in step $7$ of \blame, and still pass the PoK because of their knowledge of $\tilde{p}_k$. However, this has no benefit for the adversary, as mismatching $Q_i$'s are simply checked in the following steps of \blame, with no effect on honest parties (Property~\ref{thm10}).
\fi

	\section{Evaluation}
\label{sec:evaluation}

\noindent We implemented \prifi in Go~\citep{prifi-github}. We evaluate it on a LAN topology typical of a small organizational network.

\para{Methodology.} Our evaluation is five-fold. First, we measure the end-to-end latency via a SOCKS tunnel without data, by having a client randomly ping the relay.
Second, we compare \prifi with prior DC-nets.
Third, we replay network traces representing realistic workloads on \prifi, and measure the added latency and bandwidth usage.
Fourth, we explore limits of the system by evaluating two alternative deployment scenarios:
having a local trusted \server, which will be relevant in the case of the ICRC, and having clients outside of the LAN\@.
Finally, we explore the effect of churn and user mobility on \prifi.

\para{Experimental Setup.} We use Deterlab~\citep{deterlab} as a testbed. The experimental topology consists of a $100$Mbps LAN with $10$ms latency between the relay and the clients. We run three \servers, each having a $10$Mbps link with $100$ms latency to the relay.
We use nine machines, one dedicated to the relay and one per \server. The clients are simulated on the remaining five machines, distributed equally.
All machines are  $3$GHz Xeon Dual Core with $2$GB of RAM\@.
We focus our evaluation between $2$ and $100$ users,
which is inspired by the ICRC's operational sites (Figure~\ref{fig:icrc-cdf}).

\para{Security Parameters.} We rely on the Kyber cryptographic library~\citep{kyber}. We use $\lambda=256$~bits, Curve25519 for $\mathbb{G}$, SHA-256 as a hash function, and Schnorr signatures. The DC-net PRG uses BLAKE2 as an extensible output function.

\para{Reproducibility.} All experiments presented in this paper are reproducible with a few simple commands after cloning the repository~\citep{prifi-github}.
All raw logs and scripts to recreate the plots are available in a separate repository~\citep{prifi-github-logs}, with the exception of the private ICRC dataset.

\subsection{End-to-End Latency without Data}
\label{subsec:perfeval}

\begin{figure*}
    \vs{-0.5cm}
        \centering
        \begin{minipage}{.47\textwidth}
            \subfigure[End-to-end latency experienced by any client.
            	The baseline is twice the latency of the LAN ($20$~ms).]
            {
                \centering
                \includegraphics[width=0.9\linewidth]{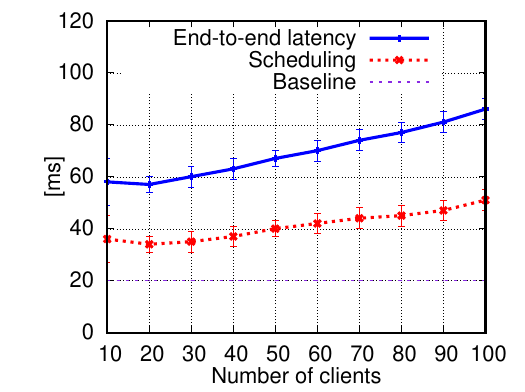}
                \label{fig:latency-inner}
            }
        \end{minipage}\hspace{.06\textwidth}\begin{minipage}{.47\textwidth}
        	\centering
            \subfigure[Latency comparison between \prifi, D\# and Riffle, computed as the time to send and decode an anonymous message.
            	The baseline is the latency of the LAN ($10$~ms).]
            {
                \centering
                \includegraphics[width=0.9\linewidth]{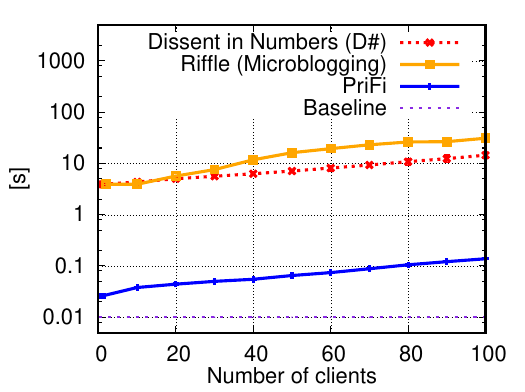}
                \label{fig:dissent}
            }
        \end{minipage}\hfill
    	\vs{-1.5em}
        \caption{End-to-end latency and comparison with the related work.}
        \label{fig:latency}
        \vs{-0.5cm}
    \end{figure*}

Figure~\ref{fig:latency-inner} shows the latency of the \prifi system, \ie the time needed for an anonymized packet to be sent by the client, decoded by the relay, and sent back to this same client.
In this experiment, one random user is responsible for measuring these ``pings'', whereas others only participate in the protocol without sending data (\ie the number of active users is $1$, anonymous among all users).
We observe that the latency increases linearly with the number of clients, from $58$ms for $20$ users (\eg a small company) to $86$ms for $100$ users, and it scales reasonably well with the number of clients. We also observe that a major component of the latency is the buffering of messages by the clients; with only one slot per schedule, clients must wait for this slot before transmitting data.

\ifextended
This waiting time is depicted by the red curve in Figure~\ref{fig:latency-inner}.
To reduce the time spent waiting on the slot, we alter the scheduling mechanism and let clients ``close'' their slot if they have nothing to send, thus enabling other clients to transmit more frequently.
This reservation mechanism improves the situation where many users are idle.
It introduces additional delay in some cases, as the client needs to wait for the next reservation to open his slot, and wait again for his slot.
Other scheduling mechanisms (\eg a compact schedule at every message, or no schedule but allowing for collisions) would yield different trade-offs between latency and number of users depending on their workload. 
\fi

\para{Pipelining.} To reduce further latency, we \emph{pipeline} rounds: we run multiple DC-net rounds in parallel instead of the ``ping-pong'' presented in \anonymize.
This enables us to better utilize the available bandwidth and reduce latency, until the capacity of the links is reached.
In this experiment, this further divides the latency by $2.25$ (Figure~\ref{fig:window}).

\para{Pipelining and Equivocation Protection.} These two components have a subtle interaction: at any point in time, the equivocation protection is computed with all the \emph{received} data, naturally ignoring ``in-flight'' data from the relay to the clients. Pipelining increases the amount of in-flight data. Importantly, this is done without packet reordering. Each message sent by the clients depends on the same received rounds, benefiting correctly from equivocation-protection.

\para{CPU\@.} Finally, during the same experiment, we briefly evaluate the CPU and memory cost on the relay (Figure~\ref{fig:cpu}).

\subsection{Comparison with Prior DC-Net Designs}

\para{Benchmarking.} We select two related works for comparison.
The closest is Dissent in Numbers (abbreviated D$\#$)~\citep{wolinsky2012dissent}. Like \prifi, D$\#$ provides provable traffic-analysis by using binary-string-based DC-nets, has similar assumptions ($M$ anytrust servers) but with no particular emphasis on being low-latency. We then compare with a more recent ACN, Riffle~\citep{kwon16riffle}: it has a similar topology but emphasizes on minimizing the download bandwidth for clients.
We do not compare with mix-nets and onion-routing protocols, whose architecture is significantly different, in that users' messages are routed \emph{sequentially} through multiple hops over the Internet.

The first major difference between \prifi and both Riffle and D$\#$ is in the functionality of the \servers: Both protocols require several server-to-server communications per round before outputting any anonymized data.

We deploy Riffle and D$\#$ on our setup and compare their latency against \prifi in Figure~\ref{fig:dissent}.
For $100$ users, a round-trip message takes $\approx14.5$~s in D$\#$ (excluding setup), $\approx31$ seconds in Riffle (which includes a one-time setup cost of $\approx7.3$ seconds), and $137$ms in \prifi (excluding setup).

\para{Higher-Level Comparison.} 
We also numerically compare \prifi against Riposte~\citep{corrigangibbs2014riposte}, a PIR-based protocol, and DiceMix~\citep{ruffing2017p2p}, a group-arithmetic-based DC-net.

Riposte uses expensive cryptography to save bandwidth.
As a result, Riposte can process up to 2 messages/sec with 2 servers (Fig. 7, p.16 of~\citep{corrigangibbs2014riposte}), whereas our relay outputs hundreds of messages/sec (Figure~\ref{fig:latency-inner}).
The topology of Riffle is precisely what we avoid in \prifi. In Riffle, the anytrust servers sequentially decrypt the client ciphertexts, then the last server broadcasts the results to all servers before any client can download it, thus increasing latency on the critical transmission path
(Figure~\ref{fig:network_setup}).

To keep a low-latency, \prifi does not use group-arithmetic-based DC-nets like DiceMix~\citep{ruffing2017p2p}.
DiceMix's latency is in the order of seconds ($\approx20$s for $100$ participants; Fig. 4 p.10 of~\citep{ruffing2017p2p}).
Although these DC-nets allow for better collision resistance and proofs of correct computation, when we evaluated them in \prifi, we found out that the generation of pseudo-randomness alone was too slow for low-latency communications.
This problem was also highlighted in Fig. 6, p.12 of Verdict~\citep{corrigan2013verdict}.

\subsection{Latency with Recorded Traffic Datasets}

\para{CRAWDAD Traces.} We evaluate the performance of \prifi when replaying the dataset `apptrafictraces'~\citep{iitkgp-apptraffic-20151126} from CRAWDAD~\citep{crawdad}.
\ifextended
We selected three sub-traces: a `Skype' trace where one client performs a VoIP (non-video) call for $281$ seconds, a `Hangouts' where one client performs a video call for $720$ seconds, and an `Others' trace where Gmail, Facebook, WhatsApp, Dropbox and other non-audio, non-video services run for $15$ minutes. 
\fi

Using the same setup as before, $5\%$ of the clients are randomly assigned packet traces from a pool and, after a random delay $r\in[0,30]$~seconds, they replay the packets through \prifi.
The relay decodes the packets and records the time difference between the decoded packet and the original trace. Because most endpoints in the traces were not reachable anymore on today's Internet,
the recorded latency does not include the communication to the Internet endpoints, but only the latency added by \prifi.

\para{ICRC Traces.} We also replay a dataset recorded at a ICRC delegation from June to July 2018.
The capture contains network-level packet headers only, corresponding to all network traffic for $30$ days of capture.
During active periods, corresponding to work days, the mean number of users is $60.9$, with a standard deviation of $5.8$.
Also during active periods, the mean bitrate of the network is of $3.1$~Mbps, with a standard deviation of $4.7$ and a (single) peak at $25$~Mbps corresponding to a bulk file transfer.
To evaluate \prifi with this dataset, we first randomly select $10$ $1$-hour periods from the active periods (\ie we exclude weekends and nights); we replay those $10$ sub-traces and measure the latency and bandwidth overhead.
During this hour, we simulate a varying number of clients: First, we identify (and only simulate) local clients, identified by an IP address of the form $10.128.10.x$; these clients replay their own packets.
Second, when needed, we add additional clients who represent extra idle users. These clients send no payload data but increase the anonymity set size.
We average the results over the $10$ $1$-hour periods and over all clients.

\begin{figure*}
	\vs{-0.5cm}
	\centering
	\begin{minipage}{.47\textwidth}
		\subfigure[Latency increase when using \prifi.
					The baseline is the latency of the LAN ($10$~ms).]
		{
			\centering
			\includegraphics[width=0.9\linewidth]{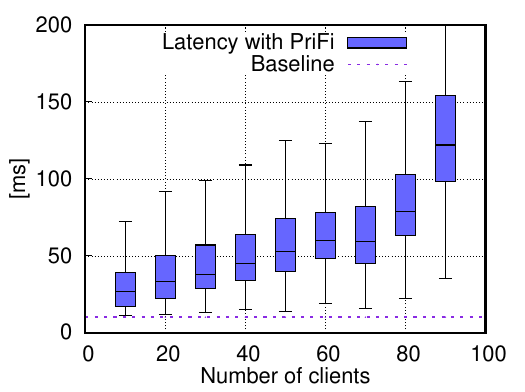}
			\label{fig:skype-lat}
		}       
	\end{minipage}\hspace{.06\textwidth}\begin{minipage}{.47\textwidth}
		\subfigure[\prifi bandwidth usage. The total bandwidth usage is the sum of last two $2$ bars (the payload is included in the LAN traffic).
		The available bandwidth in the LAN is $100$~Mbps.]
		{
			\centering
			\includegraphics[width=0.9\linewidth]{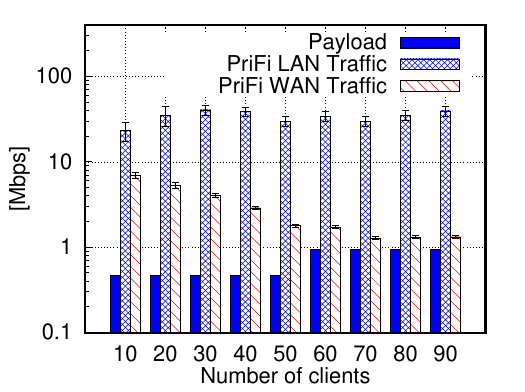}
			\label{fig:skype-bw}
		}
	\end{minipage}\hfill
	\vs{-1em}
	\caption{\prifi performance when $5\%$ of the users perform a Skype call.
				The remaining 95\% of the users are idle.}
	\label{fig:skype}
    \vs{-0.5cm}
\end{figure*}

\begin{figure*}[t]
	\centering
	\begin{minipage}{.47\textwidth}
        \subfigure[$5\%$ of users performing a Google Hangout video call. The latency drastically increases at $80$ clients, indicating the limits of the current implementation.]
        {
            \centering
            \includegraphics[width=0.9\linewidth]{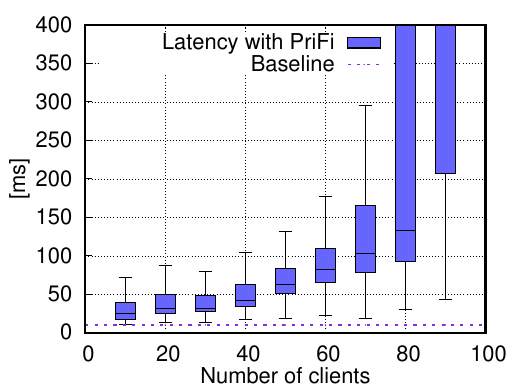}
            \label{fig:hangouts-lat}
        }
	\end{minipage}\hspace{.06\textwidth}\begin{minipage}{.47\textwidth}
		\subfigure[Latency overhead when replaying the `ICRC' dataset, with $100\%$ of users having realistic activity.
        ]
		{
			\centering
            \includegraphics[width=0.9\linewidth]{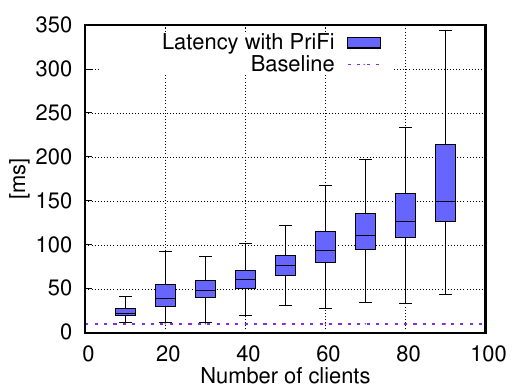}
            \label{fig:icrc-lat}
		}
	\end{minipage}\hfill
    \vs{-0.3cm}
	\caption{\prifi latency with the dataset `Hangouts' and `ICRC'.
			In both cases, the bandwidth usage (not shown here) is similar to the one observed in Figure~\ref{fig:skype-bw} (except for the payload).
			In both cases, the baseline is the latency of the LAN ($10$~ms).}
	\label{fig:hangouts-icrc}
    \vs{-0.5cm}
\end{figure*}

\begin{figure*}[t]
    \vs{-0.6cm}
	\centering
	\begin{minipage}{.47\textwidth}
        \subfigure[Larger-scale scalability. This experiment's purpose is to understand the limits of the system.]
        {
            \centering
            \includegraphics[width=0.9\linewidth]{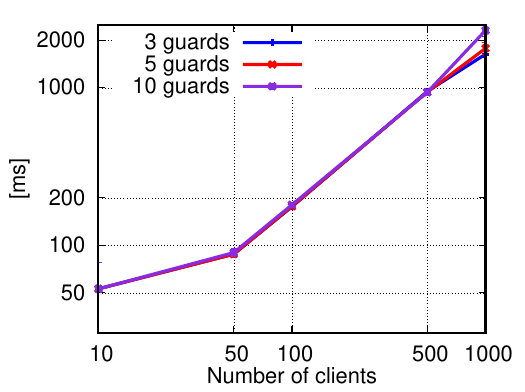}
            \label{fig:scalability}
        }
	\end{minipage}\hspace{.06\textwidth}\begin{minipage}{.47\textwidth}
        \subfigure[Latencies in the three scenarios.
        The baseline for the VPN scenario is $200$~ms, and $20$~ms in the two other scenarios.]
        {
            \centering
            \includegraphics[width=0.9\linewidth]{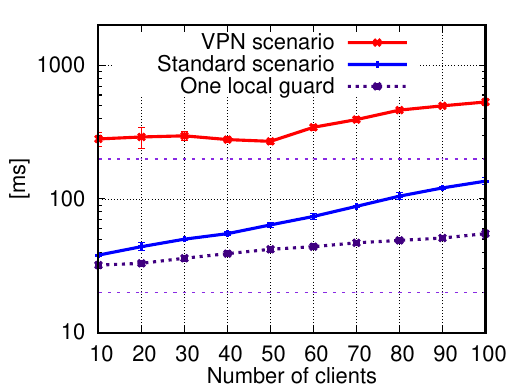}
            \label{fig:general-vs-icrc}
        }
	\end{minipage}\hfill
	\vs{-1em}
    \caption{Scalability and performance in various scenarios.}
    \vs{-0.5cm}
\end{figure*}

\ifextended

\begin{figure*}[t]
    \centering
    \vs{-0.3cm}
    \begin{minipage}{.47\textwidth}
        \subfigure[Latencies when varying the loss rate.]
        {
            \centering
            \includegraphics[width=0.9\linewidth]{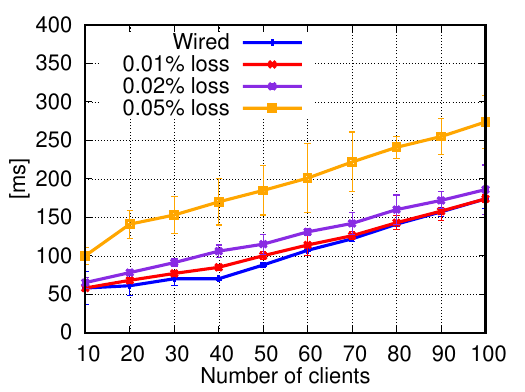}
            \label{fig:lossRates}
        }
    \end{minipage}\hspace{.06\textwidth}\begin{minipage}{.47\textwidth}
        \subfigure[Size of the anonymity set in the café scenario. This shows among how many users a \prifi client is anonymous.]
        {
            \centering
            \includegraphics[width=0.9\linewidth]{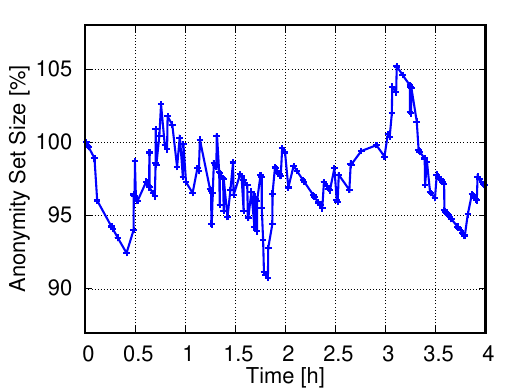}
            \label{fig:anonSetSize}
        }
    \end{minipage}\hfill
    \vs{-1em}
    \caption{Performance with varying loss rate and analysis of the impact of mobility on the anonymity set size.}
    \vs{-0.5cm}
\end{figure*}

\begin{figure}[ht!]
    \centering
    \includegraphics[width=0.9\linewidth]{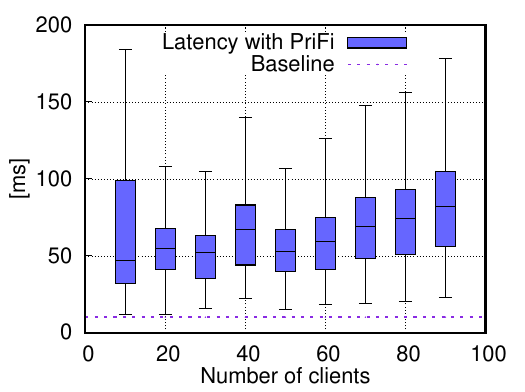}
    \caption{$5\%$ of users performing various HTTP(S) requests and file downloads.}
    \label{fig:others-lat}
\end{figure}
\fi

\para{Analysis.}
\ifextended
Figures~\ref{fig:skype-lat}, \ref{fig:hangouts-lat} and~\ref{fig:others-lat} show the added latency on the `Skype', `Hangouts' and `Other' dataset, respectively.
\fi
Figure~\ref{fig:skype-bw} shows the bandwidth used; \prifi has similar bandwidth usage for all three datasets.
The latency increases with the number of clients, which is due to (1) the increasing traffic load going through \prifi, as more users send data, and (2) to the increasing time needed to collect all clients' ciphertexts. 

In Figure~\ref{fig:skype-bw}, we show the bandwidth used by the system, split into two components: the bandwidth used in the LAN, and in the WAN\@.
We also show the bitrate of the payload, as an indication of the useful throughput (goodput) of the system.
We see that the LAN bandwidth usage is typically around $40$~Mbps, whereas the WAN usage varies from $6.9$ to $1.3$~Mbps.
We recall that, in an organizational setting, the bandwidth of the LAN is typically $100$~Mbps or $1$~Gbps, and that the bottleneck typically lies on the link towards the Internet.

In Figure~\ref{fig:skype-bw}, we see that the WAN bandwidth usage decreases with the number of clients.
This is a shortcoming that indicates that \prifi spends more time waiting and less time transmitting, due to the increased time needed to collect ciphers from more clients.
This means that \prifi cannot fully utilize the available bandwidth to offer minimal latency. This could be mitigated by increasing the pipelining of rounds for slow clients so that all clients answer in a timely fashion.
 
We learn the following: First, the mean added latency in the case of a Skype call (with $5\%$ active users) is below $100$ms for up to $80$ clients, and below $150$ms for $100$ clients.
The International Telecommunication Union estimates that the call quality starts degrading after a $150$ms one-way latency increase~\citep{itutg114}, and users start noticing a degraded quality after a $250$ms one-way latency increase~\citep{voip-info}.
Hence, the current implementation supports VoIP calls for 0--80 users and reaches its limits at around $100$ clients.
\ifextended{}
Second, the mean added latency in the case of the `Others' dataset is always below $100$ms, which seems acceptable for Web and background services; for a rough estimate, website pages usually load in seconds.
\fi
Second, the replay of the `Hangouts' data exhibits similar behavior as the `Skype' dataset; we see that the latency increases reasonably until $70$ users, but then drastically increases: the current implementation cannot transmit the data fast enough and buffering occurs at the clients.

When replaying the ICRC traces, shown in Figure~\ref{fig:icrc-lat}, we observe that the added latency varies between $15$ and $147$~ms.
This experiment was conducted with clients having network traffic corresponding to the real workload of the ICRC network. In addition, extra idle clients were simulated to achieve a constant anonymity set size of $100$.
This result confirms that \prifi can handle a realistic workload in the case of an ICRC delegation.
We emphasize that all traffic has been anonymized through the same \prifi network, regardless of QoS.
In practice, large file transfer (\eg backups) would probably either be excluded from a low-latency network, or anonymized through other means (\eg \prifi configured to provide higher throughput at higher latency).

\subsection{Scalability \& Different Scenarios}

\para{Scalability.} Figure~\ref{fig:scalability} shows the performance for larger anonymity sets and with more \servers. While the number of guards has almost no impact on performance due to the buffering at the relay, our implementation would not be low-latency for more than a few hundred clients.

\para{Local Trusted \Server.} The ICRC benefits from the particular situation of \emph{Privileges and Immunities} (P\&I)~\cite{icrc-privileges-immunities}, legal agreements with governments that provide a layer of defense in order to operate in environments of armed conflicts and other situations of violence.
In practice, P\&I notably grants the delegations with inviolability of premises and assets. Together with the strong physical security deployed at their server rooms, each delegation essentially has a local trusted server.
Aside from the ICRC, P\&I can apply to embassies and diplomatic missions~\cite{priviledges_and_immunities}.

We simulate this new deployment with one \server in the LAN instead of three remote \servers.
The latency between the relay and the unique \server is $10$~ms.
In this case, we observe that the latency experienced by clients is roughly cut in half, shown in Figure~\ref{fig:general-vs-icrc}, purple dotted curve versus blue solid curve.
An additional benefit is that the only WAN bandwidth usage is the anonymized goodput.

\para{VPN.} When a member of an organization is accessing the network remotely (\eg when traveling), it can benefit from \prifi's protection from outside the organizational LAN\@.
The cost is in performance, as the relay waits upon the slowest client to decode an anonymous message.

We simulate this alternative deployment scenario by having \emph{all} clients outside the LAN\@. 
This is modeled by setting the latency between clients and relay to $100$~ms instead of $10$~ms.
In this case, the baseline for latency is $200$~ms.
We observe that in this scenario, end-to-end latency varies from $280$~ms to $498$~ms as shown in Figure~\ref{fig:general-vs-icrc}, red solid curve versus blue solid curve.
While this latency is too high to support VoIP and videoconferencing, it might be acceptable for web browsing.
We note that all users are slowed down. Although not explored in this work, this slowdown could be mitigated by having two \prifi networks, one reserved for local users, the other accepting remote users.
Local users would participate in both networks, ensuring a sufficiently large anonymity set, and would communicate only using the fastest \prifi network.

\para{Loss Rates.} We briefly explore the impact of various loss rates in Figure~\ref{fig:lossRates}. While an imperfect representation,
this experiment could sketch the performance in a real WLAN\@.
The results show that the current implementation requires ``good'' link quality  (loss rates $\le10\%$, see~\citep{sheshadri2017packet} Figure 3a) to maintain low latency,
that then degrades rather quickly with increasing loss rates.
We note that current WLANs typically have good resilience to message drops; noise and collisions result in increased jitter rather than losses~\citep{korhonen2005effect}.
Implementing \prifi directly on Network Interface Cards (NIC) could give better control and performance.
Finally, we note that WLANs have a less expensive broadcast than LANs, a factor not reflected in this experiment.
\subsection{Client Churn}
\label{subsec:eval-client-churn}

	In DC-nets, churn invalidates the current communications and leads to data re-transmissions and global downtime where no one can communicate.
Although re-transmissions are acceptable with \prifi's small and frequent rounds (\eg a few $100$KB of payload each $10$ms), frequent churn could prevent delay-sensitive applications from running on top of \prifi.
To our knowledge, our contribution here is the first analysis of the impact of churn on DC-nets in a realistic scenario where nodes are mobile (\eg~Wireless devices).

\para{Dataset.} To characterize node mobility, we use a standard dataset~\citep{pdx-vwave-20070914} from CRAWDAD~\citep{crawdad}.
It contains four hours of wireless traffic, recorded in a university cafeteria. Those traces contain the Data Link layer (and show the devices' association and disassociation requests).
The dataset contains $254$ occurrences of churn over $240$ minutes, in which there are $222$ associations ($33$ unique devices) and $32$ disassociations ($12$ unique devices).
In comparison with the ICRC scenario, this dataset is likely a pessimistic model, as node mobility in a cafeteria is likely higher than in offices.

\para{Dataset Analysis.} Each device (dis)connection induces a re-synchronization time of $D$ milliseconds (for \setup), where $D$ is dominated by the number of \servers $m$ and clients $n$ and the latency between them; a typical value for $D$ would be a few seconds (Appendix, Figure~\ref{fig:downtime}).

\vs{0.2em}
\noindent We analyze two strategies to handle churn: 

\squishlist
    \item The \emph{naïve} approach stops communication for $D$ sec at every churn.
    \item The \emph{abrupt} disconnections stops communication for $D$ sec at every disconnection only. Devices connect using a graceful approach (\setup done in the background, keeping the previous \anonymize protocol running until the new set of clients is ready).
    	This strategy can be enforced by the relay.
\squishend

In Table~\ref{table:churn}, we show (1) the raw number of communication interruptions, which directly comes from the node mobility in the dataset, (2) the network availability percentage, computed as $\frac{1-\text{downtime}}{\text{total time}}$, and (3) the maximum continuous downtime, \ie the longest network interruption.

\para{Analysis.} Using \prifi in the cafeteria scenario represented by the dataset~\cite{pdx-vwave-20070914} would naturally decrease network availability, as churn induces global downtime.
Over $4$h, between $0$ and $32$ global losses of communication occur, and the network availability ranges between $99.82\%$ and $100\%$, depending on the disconnection types.
The longest disconnection period is $0.82$s in the worst case, which might be noticeable as a slight lag by users of time-sensitive applications (\eg VoIP). If we extrapolate the worst-case availability $99.82\%$ over $10$~minutes, the total expected downtime is below $1.1$~second. We note that the metric does not directly compare with the classical ``five 9's'' in Service Level Agreements, for which in the worst case all downtime can be continuous (\eg a server is down for a few minutes); in our model, downtime is likely to be spread over many short events over time, following user mobility.

\begin{table}[t]
	\caption{Effect of different churn handling strategies on \prifi.}
	\begin{tabular}{llll}
		\textbf{Strategy} & \textbf{\# Interrupt.} & \textbf{Availability [\%]} & \textbf{Max Downtime [s]} \\
		Naïve             & 254                    & 98.72792                   & 1.55147                   \\
		Abrupt            & 32                     & 99.81778                   & 0.82                      \\
	\end{tabular}
	\label{table:churn}
\end{table}

\begin{figure}[t]
    \centering
    \includegraphics[width=0.8\linewidth]   {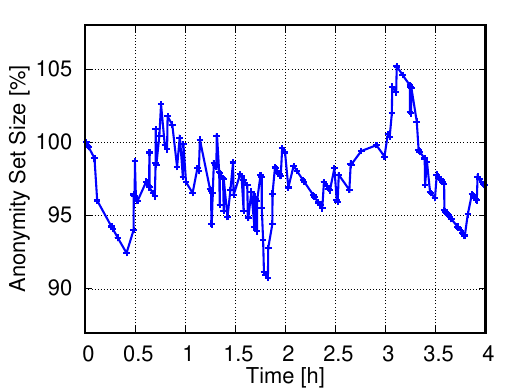}
    \caption{Size of the anonymity set versus time, when using \prifi in the café scenario. This shows among how many users a \prifi client is anonymous.}
    \label{fig:anonSetSize2}
    \vs{-1.5em}
\end{figure}
    
\para{Anonymity Metrics.} We now analyze the size of the anonymity set with respect to time, \ie among how many participants a \prifi user is anonymous at any point in time (Figure~\ref{fig:anonSetSize2}).
In particular, we are interested in the variations (in this case which are due to user mobility).
A high variance means that while connected, a user risks being ``less'' anonymous if unlucky (and many people disconnect suddenly); if the size of the anonymity set drops to $1$, anonymity would be lost.
We display the difference, in percentage, between the actual anonymity set size and the baseline tendency. We see that size of the anonymity set does not vary more than $\rpm8\%$, and the mean number of users is $50$.
Hence, our estimation for the worst-case of ``anonymity loss'' in this scenario is of $4$ users, which seems acceptable in an anonymity set of $50$ users.

\section{Discussion on Intersection Attacks}
\label{sec:intersections}
Goal G1 ensures that a client's message is anonymous among other honest participating members.
In the case where a client performs the same recognizable action (e.g., contacting a particular website) in two different epochs, 
the adversary can guess that this anonymous client is the same in the two epochs, and the anonymity set of this client is reduced. 
This problem can be exacerbated if these recognizable actions are performed over a long period of time. Also, to facilitate intersection attacks, the malicious relay could actively kick out clients.
\prifi does not have a perfect solution to the problem, but we suggest the following approaches.

First, in the context of an organization, desktop computers should be connected to \prifi most of the time, providing a baseline anonymity set.

Second, as is the case with Tor, users should ideally be cautious of ``blending in'' by having traffic patterns similar to other users.
This could be partially enforced by the organization by having a standard set of applications.
If the destinations are sensitive (\eg a specific website regularly visited by one client), all clients could periodically make requests to random recently-contacted websites, for which a public, signed list would be broadcast by the relay. Incidentally, this would improve deniability \emph{and} accountability: knowing that the list of destinations is public, an employee is less inclined to misbehave.

More importantly, clients should communicate only when a sufficiently large number of users are online, for example by using a system such as Buddies~\citep{wolinsky2013hang} and communicating only when their trusted ``friends'' are online.
Considering that in an organization's building, most people roughly share the same working hours, we postulate that this requirement is not too constraining.

We note that traditional approaches such as anonymous authentication protocols (\eg DAGA~\citep{syta14deniable}) seem fundamentally insufficient, as the clients are potentially directly connected to the malicious relay, which could actively fingerprint them using a plethora of techniques~\citep{desmond2008identifying,hall2004enhancing,franklin2006passive}. In this case, a segregated network topology could make the task harder for the adversary, but could increase the cost of broadcasting downstream messages.

    \section{Related Work}
\label{sec:related-work}

One straightforward way to protect against local eavesdroppers is by tunneling the traffic through a VPN that is outside of the adversary's control. 
However, this provides no guarantee when the VPN provider is malicious. Moreover, VPNs protect neither communication patterns nor, in the case of a local eavesdropper, the communication source.

\vs{0.1cm}
\noindent Anonymous Communication Networks (ACN) are the closest related work, but existing solutions translate poorly to the LAN setting:

\para{Onion Routing.} Tor~\citep{dingledine2004tor} 
does not provide traffic analysis resistance against a global passive adversary~\citep{pfitzmann95breaking, nguyen03breaking, panchenko2011website, wang2013improved, wang2014effective}.
Some prior work focuses on low latency and efficiency,
at the cost of traffic-analysis resistance (LAP~\citep{hsiao2012lap}, Dovetail~\citep{sankey2014dovetail}, Hornet~\citep{chen2015hornet}).
Taranet provides traffic-analysis resistance through traffic shaping~\citep{chen2018taranet}. Ricochet~\citep{ricochet} and Pond~\citep{pond} are messaging systems that build on Tor.

\para{Mix Networks.} Older solutions such as Crowds~\citep{reiter1998crowds}, Mixminion~\citep{danezis2003mixminion}, and Tarzan~\citep{freedman2002tarzan} are not traffic-analysis resistant.
Although more recent work addresses this issue,
due in part to the Anonymity Trilemma~\citep{das2018anonymity},
this incurs high costs in either latency or bandwidth.
A common way to control these costs is by having application-\emph{specific} ACNs:
Vuvuzela~\citep{van2015vuvuzela}, Atom~\citep{kwon2017atom}, Karaoke~\citep{lazar2018karaoke}, Loopix~\citep{piotrowska2017loopix}, Stadium~\citep{tyagi2017stadium}, XRD~\citep{kwon2020xrd} for messaging/microblogging/tweets; Herd~\citep{le2015herd}, Yodel~\citep{lazar2019yodel} for VoIP\@; Herbivore~\citep{goel2003herbivore}, Aqua~\citep{le2013towards} for file sharing.

\vspace{0.1cm}

\noindent Both onion-routing and mix-networks are sub-optimal in providing anonymity in an organizational network.
Servers should typically be non-colluding, and having them all in the same LAN reduces collective trustworthiness.
When the servers are outside the LAN, both designs route users' traffic over the Internet, adding latency.

\para{Differential Privacy.} Differential privacy~\citep{dwork2014algorithmic} allows leaking a small amount of statistical
information about the user's communication. These systems are typically built on mixnets~\citep{van2015vuvuzela,lazar2016alpenhorn,tyagi2017stadium,lazar2018karaoke}.

\para{PIR.} A category of ACN relies on PIR~\citep{chor1995private} and ORAM~\citep{boneh2011remote} to implement efficient messaging systems (Riposte~\citep{corrigangibbs2014riposte}, Riffle~\citep{kwon16riffle}, Pung~\citep{angel2016unobservable}). 
These are typically not made for low latency, as the anonymity set is built over a time period.

\para{SDN.} Some solutions use Software Defined Networks to anonymize packet headers, but they are typically not traffic-analysis resistant against active attacks~\citep{skowyra2016have,meier2017itap,zhu2016mic}.

\para{DC-nets.} DC-nets~\citep{chaum1988dining, corrigan2010dissent,wolinsky2012dissent} have provable traffic-analysis resistance and typically have high bandwidth and latency costs.
\prifi fits this category and focuses on low latency in the context of organizational networks.
\noindent Unlike the binary-string-based DC-nets used in \prifi, some related works rely on group arithmetic (Verdict~\citep{corrigan2013verdict}, \citep{golle2004dining}, DiceMix~\citep{ruffing2017p2p}), which enables computations to be proven correct. Unfortunately, their high computational cost makes them unsuited for low-latency applications.

DC-nets have been further studied with an emphasis on collision resolutions~\citep{garcia2015storage, garcia2014beating} and user scheduling~\citep{krasnova2016footprint}.

\para{PriFi16.} \prifi builds upon PriFi16~\citep{barman2016prifi},
that sketched the idea of DC-nets for LANs. 
Despite considering malicious insiders, PriFi16 does not provide a solution to insider jamming. 
PriFi16 acknowledges the problem of the equivocation without providing a concrete solution.

\para{Summary.} \prifi is a low-latency, traffic-agnostic solution working like a VPN, conceptually close to Tor~\citep{dingledine2004tor}, Hornet~\citep{chen2015hornet} and Taranet~\citep{chen2018taranet}, but tailored for LANs/WLANs.


	\section{Conclusion}
\label{sec:conclusion}

We have presented \prifi, an anonymous communication network that protects organizational networks from eavesdropping and tracking attacks. \prifi exploits the characteristics of (W)LANs to provide low-latency, traffic-agnostic communication. 

\prifi reduces the high communication latency of prior work via a new client/relay/server architecture. This new architecture removes costly server-to-server communications, and allows client's traffic to be decrypted locally, remaining on its usual network path. This avoids the latency bottleneck typically seen in other systems.

\prifi also addresses two shortcomings of the related work: First, users are protected against equivocation attacks without added latency or costly gossiping; second, leveraging the LAN topology, disruption attacks are detected retroactively and orders of magnitude faster.


We have implemented \prifi and evaluated its performance on a realistic setup, mimicking the targeted ICRC deployment.
Our findings show that various workloads can be handled by \prifi, including VoIP and videoconferencing, and that restrictions usually imposed by DC-nets in case of churn when users are mobile are not problematic in \prifi.

\para{Acknowledgments.} We are thankful to Vincent Graf Narbel, Alejandro Cuevas, Caleb Malchik, Jun Chen, Lucas Gauchoux, Matthieu Girod, Pierre Sarton, Yannick Schaeffer and Julien Weber for their valuable help and feedback on this project.

This research was supported in part by U.S. National Science Foundation grants CNS-1407454 and CNS-1409599, U.S. Department of Homeland Security grant FA8750-16-2-0034, U.S. Office of Naval Research grants N00014-18-1-2743 and N00014-19-1-2361, and by the AXA Research Fund.

    \newpage
	
	{
        \vs{-0.6cm}
        \section*{References}
        \vs{-0.3cm}
        \renewcommand{\section}[2]{}%
		\bibliographystyle{abbrvnat}
		\bibliography{bibliography.bib}
	}

\appendix
\counterwithin{figure}{section}

\vs{-0.6cm}
\section{Proofs of Properties}
\label{sec:proofs}
\vs{-0.3cm}

\subsection{Setup}
\label{subsec:proofs-setup}
\vs{-0.3cm}

\begin{theorem}\label{thm1}
	A shared secret $r_{ij}$ between an honest client $C_i$ and an honest \server $S_j$ is known only to $C_i$ and $S_j$.
\end{theorem}

\begin{proof}
	The shared secrets $r_{ij}$ are derived using an authenticated Diffie-Hellman protocol; due to the hardness of the DDH assumption in $\mathbb{G}$ (which implies hardness of DLP), $\mathcal{A}$ is unable to recompute $r_{ij}$ without the private key of $C_i$ or $S_j$.
\end{proof}

\begin{theorem}\label{thm2}
	Let $C_0$, $C_1$ be two honest clients who ran \setup without aborting, and $\alpha(0), \alpha(1)$ the position their respective shuffled key in $\pi$.
	Then, the adversary $\ad$ has negligible advantage in guessing $b\in[0,1]$ with significant advantage such that the mappings client $\to$ position $(b\to \alpha(0), (1-b)\to \alpha(1))$ are in $\pi$. 
\end{theorem}

\begin{proof}	
	If \setup terminates without aborting for an honest client $C_i$, then, as a consequence of the checks done by $C_i$, it means that: (1) the verifiable shuffle completed correctly, (2) $\pi$ is signed by every \server in $G$, (3) there are at least $K=2$ clients in $\pi$, and (4) their own pseudonym is included in $\pi$.
	
	Without loss of generality, let us assume that $S_1$ is the honest server in $G$. Since $\pi$ is signed by all $S_j\in G$, then $S_1$ participated in Neff's Verifiable Shuffle~\citep{neff2003verifiable} and signed the output $\pi$.
	
	Since $G$ contains at least $2$ honest users, and both $C_1$ and $C_2$ ran \setup without aborting, it means that their two pseudonyms are included in $\pi$. Since both are honest, their ephemeral private keys $p_1,p_2$ are unknown to $\ad$.
	
	Therefore, the Neff shuffle ran with at least $2$ keys as input, which are the keys of the honest clients (due to check 4), and at least one honest server shuffled the keys. Let $\alpha(0), \alpha(1)$ be the position of $C_1$ and $C_2$'s respective shuffled keys in $\pi$. Without $p_1$, $\ad$ is unable to differentiate between $\tilde{P}_{\alpha(0)}$ and $\tilde{P}_{\alpha(1)}$, and can do no better than random guessing. The same argument can be made for $C_2$.
\end{proof}

\vs{-0.3cm}
\subsection{Anonymize}
\label{subsec:proofs-anonymize}
\vs{-0.3cm}

\begin{theorem}\label{thm3}
	After a run of \texttt{Anonymize}, let $C_{i_1},C_{i_2}$ be two honest clients, $k_1=\alpha(i_1), k_2=\alpha(i_2)$ be the time slots in which they communicated, and let $m_{k_1},m_{k_2}$ be the anonymous upstream messages for those slots.
	Then, $\mathcal{A}$ has negligible advantage in guessing $b\in[1,2]$ such that $m_{k_b}$ is the message sent by $i_1$.
\end{theorem}

\begin{proof} \noindent $\mathcal{A}$ tries to distinguish between $(C_{i_1}\to m_{k_1}, C_{i_2}\to m_{k_2})$ and $(C_{i_1}\to m_{k_2}, C_{i_2}\to m_{k_1})$.
	For simplicity, we define the following equivalent game: for a slot $k\in\{k_1,k_2\}$, $\mathcal{A}$ guesses $b\in[1,2]$ such that $C_{i_b}$ is the anonymous sender of the message $m_{k}$.
	
	To ease notation, let $C_{i_1}$ be $C_1$, and $C_{i_2}$ be $C_2$.
	Without loss of generality, let $S_1$ be a honest \server, and let $C_1$ be the anonymous sender.
	Then, on slot $k_1$, we have
	
	\begin{equation}
	\begin{split}
	c_1 & = \bigoplus_{j=1}^m~\texttt{PRG}(r_{1,j})~\oplus~m_1\\
	c_2 & = \bigoplus_{j=1}^m~\texttt{PRG}(r_{2,j})
	\end{split}
	\end{equation}
	
	Since $S_1,C_1$ and $C_2$ are honest, the values $\texttt{PRG}(r_{1,1})$ and $\texttt{PRG}(r_{2,1})$ are unknown to $\mathcal{A}$.
	We isolate the contribution of $S_1$:
	\begin{equation}
	\begin{split}
	c_1 & = \bigoplus_{j=2}^m~\texttt{PRG}(r_{1,j})~\oplus~\texttt{PRG}(r_{1,1})~\oplus~m_1\\
	c_2 & = \bigoplus_{j=2}^m~\texttt{PRG}(r_{2,j})~\oplus~\texttt{PRG}(r_{2,1})
	\end{split}
	\end{equation}
	
	All values have the same length.
	Both values $\texttt{PRG}(r_{1,1})$ and $\texttt{PRG}(r_{2,1})$ are the output of a PRG and are indistinguishable from a random string in the random oracle model.
	Therefore,  $\texttt{PRG}(r_{1,1})~\oplus~m_1$ is also indistinguishable from a random string, and hence so are both strings $c_1$ and $c_2$.
	Hence, since $r_{1,1}$ and $r_{1,2}$ are unknown to $\mathcal{A}$, if $\texttt{PRG}$ is correctly instantiated, $\mathcal{A}$ sees two random strings and can only guess $b$ with probability $1/2$.
\end{proof}

\vs{-0.8cm}
\subsection{Disruption-Protection}
\label{subsec:proofs-disruption}
\vs{-0.3cm}

\begin{theorem}\label{thm4}
	The anonymity of any honest client is unaffected by the information $\vec{{\texttt{PRG}(r_{ij})}_l}$ made public in step $2$ of \blame (Protocol~\ref{alg:disruption-blame}).
\end{theorem}

\begin{proof}
	The values ${\texttt{PRG}(r_{ij})}_l$ for slot $k$ de-anonymize precisely the $l^\text{th}$-bit of the slot $k$, by revealing the composition of the messages $\vec{c_i}$ and $\vec{s_j}$ at position $l$.
	
	Let $C_i$ be the (honest) owner of slot $k$.
	Only $C_i$ can generate the proof of knowledge $\text{PoK}_{k,l}(\tilde{p}_k: \tilde{p}_k=\log~\tilde{P}_k)$, and honest clients verify the $\text{PoK}$ before revealing any information, hence honest clients do not reveal information about a slot without the owner's consent.
	
	In slot $k$, all non-sender, honest clients transmitted a $0$ on bit $l$, since $k\neq \alpha(i)$ and the protocol forces them to transmit $\vec{0}$.
	Additionally, the slot owner $C_i$ only sends a $\text{PoK}$ for $l$ such that ${(m_k)}_l=0$.
	Therefore, on slot $k$, at position $l$, all honest clients transmitted a $0$.
	The values revealed match the ${(c_i)}_l$ values previously sent, and the ${(c_i)}_l$ are composed solely the output of the PRGs seeded by the secrets, which are indistinguishable from a random string in the random oracle model. Finally, the knowledge of one given bit reveals nothing about other parts of the PRG output.
\end{proof}

\begin{theorem}\label{thm5}
	The anonymity of any honest client is unaffected by the information $r_{ij}$ made public in step $5$ of \blame (Protocol~\ref{alg:disruption-blame}).
\end{theorem}

\begin{proof} By revealing $r_{ij}$, a pair of client-\server $(C_i,S_j)$ completely reveal their contributions to the DC-net.
	However, a honest client and a honest \server never contradict each other, as the protocol ensures that their values ${\texttt{PRG}(r_{ij})}_l$ are correctly computed from the common shared secret $r_{ij}$.
	Since in step $5$, a honest client $C_i$ only reveals $r_{ij}$ when seeing (1) a signed message from a \server, (2) for which there is a contradiction, he never reveals $r_{ij}$ for an honest \server $S_j$.
	The same argument can be make for $S_j$ about any honest client.
	
	Therefore, those values are revealed only when $C_i$ or $S_j$ is malicious (or both); in this case, $\mathcal{A}$ already had knowledge of $r_{ij}$.
\end{proof}

\begin{theorem}\label{thm7}
	Let $C_i$ be the owner of a slot $k$, and let $C_d,~d\neq i$ be another client (or \server).
	If $C_d$ sends an arbitrary value $c_i'$ instead of the value $c_i$ (or $s_j$) as specified in the protocol, then $C_d$ is identified as the disruptor and is excluded from subsequent communications.
\end{theorem}

\begin{proof}
	We show the following slightly weaker statement: in the case of multiple disruptors, the protocol excludes \emph{a} disruptor $C_d$.
	We argue that since all are under the control of $\mathcal{A}$, \emph{which} disruptor gets excluded is irrelevant. By repeating this argument, eventually, all disruptors gets excluded.
	
	The proof describes a client disruptor; the argument for a \server is analogous.
	
	Let $c_i'$ be the value sent instead of $c_i$.
	Then, we can write $c_i'=c_i~\oplus~q$, and we observe that $m_k = m_i~\oplus~q$.
	Let $|m_i|$ be the length of $m_i$, and $z$ be the number of $0$ bits in $m_i$.
	The adversary disrupts stealthily (\ie is not detected) if no $0$ in $m_i$ is flipped to a $1$ (in this case, the honest client cannot start a \blame without de-anonymizing himself).
	
	If $m_i$ is unknown to $\ad$ and is uniformly distributed in $\bin^{|m_i|}$, the probability that no $0$-bit is flipped follows ${(\frac{1}{2})}^z$, with $\mathbb{E}[z]=\frac{|m_i|}{2}$.
	
	If, however, $\ad$ has knowledge of some distribution of $m_i$ (\eg because perhaps, after a long period of silence, the first message of a client is often an unencrypted DNS request), then $\ad$ has a much better chance of targeting the $1$ bits.
	Fortunately, the clients are free to compose $m_i$ how they want, \eg by starting with several 0's ($0\ldots0||\text{DNS}||0\ldots0$), so we expect that in practice the adversary should have little information about the distribution of each bit in $m_i$.
	We note that clients can notice disruption and vary how they compose $m_i$, or even send a ``trap-message'' of all $0$'s (which allows to catch a disruptor with probability $1$) should they notice significant disruption.
\end{proof}

\begin{theorem}\label{thm6}
	An honest entity is never identified as a disruptor.
\end{theorem}

\begin{proof}
	Trivially, honest clients follow the protocol.
	Each step of the \blame protocol consists of revealing some already-performed step of \setup and \anonymize protocol; by definition, honest entities performed those correctly.
\end{proof}

\vs{-0.8cm}
\subsection{Equivocation-Protection}
\label{subsec:proofs-equivocation}
\vs{-0.3cm}

\begin{theorem}\label{thm8}
	The anonymity of any honest client is unaffected by the extra information $\kappa_i$ sent in \texttt{DCNet-Gen-Client}.
\end{theorem}

\begin{proof}
	For a slot $k$ which does not go through a \blame procedure, the adversary knows $\kappa_i$ for all clients, $h_i$, and all $\texttt{PRG}(r_{ij})$ for a malicious $C_i$, or a malicious $S_j$, but not between a honest pair $(C_i,S_j)$.
	Without loss of generality, assume $C_1,C_2,S_1$ are the honest clients and \server.
	Then, $\kappa_1=F_1(\gamma) + c_1 + c_2$, where $c_1$ is a constant known to $\mathcal{A}$, and $c_2=F_2(h_i)\cdot F_3(H(\texttt{PRG}(r_{1,1})))$ is unknown to $\mathcal{A}$ and distributed uniformly at random in $\mathbb{G}$ due to $F_2$.
	$\mathcal{A}$ is unable to distinguish between $F_1(\gamma)+c_2$ and $c_2$, both being random values uniformly distributed in $\mathbb{G}$, and hence cannot distinguish between the $k_i$ of a honest sender and of a honest non-sender.
	
	If the slot $k$ goes through a \blame procedure, then in addition to the previous knowledge, for any $i,j$, at most one bit of $\texttt{PRG}(r_{ij})$ is revealed in step 2 of \blame; this value is used in the computation of $\kappa_i$ and $\sigma_j$, but this single bit does not reveal information about $H(\texttt{PRG}(r_{ij}))$. 
	
	A private value $\texttt{PRG}(r_{ij})$ (between a pair of honest client/\server $C_i$ and $S_j$) cannot be recovered from $\kappa_i$ by $\mathcal{A}$ due to the use of the cryptographic hash function $H$.
\end{proof}

\begin{theorem}\label{thm9}
	The anonymity of any honest client is unaffected by the extra information $\sigma_j$ sent in \texttt{DCNet-Gen-\Server}.
\end{theorem}

\begin{proof}
	The $\sigma_j$ are independent from the communicated content and the slot owner, and reveal nothing about these. 
	
	As discussed in Proof of Property~\ref{thm8}, due to the use of the cryptographic hash function $H$, the $\sigma_j$ do not allow to recover any $\texttt{PRG}(r_{ij})$, nor does revealing $1$ bit of $\texttt{PRG}(r_{ij})$ allow to recover $H(\texttt{PRG}(r_{ij}))$.
\end{proof}

\begin{theorem}\label{thm10}
	The anonymity of any honest client is unaffected by the extra information $\vec{Q_i}, \text{PoK}$ sent in step 7 of \blame.
\end{theorem}

\begin{proof}
	The $Q_i$ are independent from the communicated content and the slot owner, and reveal nothing about these. 
	
	Each value $Q_i$ in $\vec{Q_i}$ has the form $Q=P\cdot F_2(H(\texttt{PRG}(r_{ij})))$, where $P$ is a base point of $\mathbb{G}$, in which the DLP problem is hard. 
	Without loss of generality, assume $C_1,C_2,S_1$ are the honest clients and \server.
	Then, $F_2(H(\texttt{PRG}(r_{11})))$ and $F_2(H(\texttt{PRG}(r_{21})))$ are unknown to $\ad$, and the $Q_i$ do not help in computing them due to the assumption on $\mathbb{G}$. Therefore, both $C_1$ and $C_2$ send a value (1) unknown to $\ad$ and (2) uniformly distributed in $\mathbb{G}$ due to $F_2$, hence $\ad$ has no advantage in telling them apart.
	
	Due to the hardness of the DLP in $\mathbb{G}$, the values $Q_i$ give no information about $\kappa_i$.
	
		
	Finally, the zero-knowledge proof of knowledge $\text{PoK}$ trivially reveals no information.
\end{proof}

\begin{theorem}\label{thm11}
	The anonymity of any honest client is unaffected by the information $r_{ij}$ made public in step $9$ of \blame.
\end{theorem}

\begin{proof} 
	This proof is a direct copy of the proof of Property~\ref{thm5}.
\end{proof}

\begin{theorem}\label{thm12}
	If $\exists i,j$ two honest clients who received $\vec{d}_i \neq \vec{d}_j$ on round $k$, then neither the relay, nor $\mathcal{A}$ can decrypt $m_k$ for any subsequent round $k'>k$.
\end{theorem}

\begin{proof}
	Without loss of generality, let $C_1, C_2$ be two honest clients who received $\vec{d}_1 \neq \vec{d}_2$ at the end of round $k$.
	Then, with overwhelming probability, $h_1\neq h_2$ due to the use of the cryptographic hash function, and it follows that $F_2(h_1)\neq F_2(h_2)$ with high probability. 
	
	To simplify the proof, assume $h_1$ is the ``correct'' history.
	Therefore, the correct value for $\kappa_2$ (if $C_2$ had received the correct downstream message $\vec{d}_1$) would be
	
	\[\kappa_2' = F_2(h_1)\cdot\sum_{j=1}^{m}F_3(H(\texttt{PRG}(r_{ij})))\]
	
	Without loss of generality, assume $S_1$ is the honest \server; we rewrite the above equation, isolating $r_{2,1}$:
	
	\[\kappa_2' = F_2(h_1)\cdot F_3(H(\texttt{PRG}(r_{2,1}))) + F_2(h_1)\cdot\sum_{j=2}^{m}F_3(H(\texttt{PRG}(r_{ij})))\]
	
	But since $C_2$ and $S_1$ are honest, $r_{2,1}$ is unknown to $\mathcal{A}$ and so is the scalar $F_3(H(\texttt{PRG}(r_{2,1})))$, and therefore $\mathcal{A}$ is unable to recompute the decryption key $\gamma$ for any round $k'>k$.
	
	We note that the proof does not requires $C_1$ nor anyone to have a ``correct'' history, simply $h_1 \neq h_2$ for two honest clients.
	
	Additionally, while this property ensures that an equivocation attack never leads to de-anonymizing the client through their anonymous output, incidentally, it gives honest clients a proof $\mathcal{S}_{\hat{p}_r}(h_r)$, with $h_r\neq h_i$, that the relay did not behave correctly; administrative can then be taken against the relay.
\end{proof}

\begin{theorem}\label{thm13}
	If a client $C_i$ sends an arbitrary value $\kappa_i'$ instead of the value $\kappa_i$ as specified in the protocol, then $C_i$ is identified as the disruptor and is excluded from subsequent communications.
\end{theorem}

\begin{proof}
	The argument is analogous to the proof of Property~\ref{thm7}: if the honest client whose slot was disrupted can start a blame, then the \blame protocol ultimately excludes one disruptor.
\end{proof}
	
\begin{theorem}\label{thm14}
	An honest relay is never accused of an equivocation attack.
\end{theorem}
	
\begin{proof}
	Since the since the downstream messages $\vec{d}$ are signed by the honest relay, the adversary is unable to forge a message sent by the honest relay. Since an honest relay never equivocates, a malicious client attempting to frame the relay is equivalent to a malicious client sending wrong $\vec{c_i},\vec{\kappa_i}$ values.
	
	When this happens during an honest client slot, it will eventually start a \blame procedure. The \blame procedure verifies the correctness of the $\vec{c_i},\vec{\kappa_i}$ values since, for a correct $h_i$, the $\vec{c_i},\vec{\kappa_i}$ of all parties but the honest sender are fully determined by the shared secrets $r_{ij}$. The honest sender trivially sends the correct values.
	
	When this happens during a malicious client slot, the client may wrongly compute $\kappa_i$ and still pass the \blame due to the first ``Or'' of the PoK\@. In this case, no party is identified as a disruptor.
	
\end{proof}
	
\begin{theorem}\label{thm15}
	An honest client is never excluded as a disruptor.
\end{theorem}
	
\begin{proof}
	The end of \blame (Protocol~\ref{alg:disruption-blame}) describes more precisely what it means for a client $C_i$ to be ``excluded as a disruptor'': the relay broadcasts a new group $G'$ and roster $T'$ without the public key of $C_i$. The process for a server is analogous, except that $T'=T$.
	
	The proof is done for an honest client, and is analogous for an honest server.
	
	Trivially, an honest client always produces correct values. During a \blame procedure, every message is signed. The honest client first sends $\vec{\texttt{PRG}(r_{ij})}$ for slot $k$, which matches what he sends in round $k$. Then, if he is not in contradiction with some server, the honest client stops taking part in \blame (and is not excluded); let us assume he is in contradiction with $S_j$. Then, upon reception of the signed message from $S_j$, he reveals $r_{ij}$ along with a proof of correctness. 
	
	To exclude a client (by broadcasting a new group $G'$ and roster $T$), the relay also has to reveal all inputs of \blame for accountability. To exclude an honest client without blatantly cheating, the relay would have to forge one signature, which contradicts our adversarial model.
	
	In practice, we note that the malicious relay could update $G$ and $T$ without justification; however, this is analogous to a simple denial-of-service and contradicts the threat model. In practice, this would prompt clients to take administrative actions against the relay.
\end{proof}
\ifextended{}
	\section{Supporting Different QoS Levels} 
	\label{sec:qos}
	
	To support different QoS levels, \prifi enables clients to subscribe to channels with different bitrates.
	Each channel is an isolated instance of the \prifi protocol, but runs with different speeds and payload sizes.
	In this way, constrained devices (\eg suitable IoT devices, battery-powered devices)  only join ``slow'' \prifi channels that require little computation, and more powerful devices join both slower and faster channels.
	
	Channels correspond to categories of traffic, for instance e-mails, web browsing, VoIP and video conferencing.
	With this approach, a device that has no VoIP capabilities can save resources by joining only the appropriate channel.
	
	Devices connected to several channels participate in the anonymity set of those channels, but  probably only communicate using the fastest.
	The benefit of this approach is that the size of the anonymity set increases for slower channels.
	The cost of being active in several channels is estimated as follow: assuming an order of magnitude difference in terms of latency between channels (\eg web browsing at $500$ms, VoIP at $50$ms latency), joining an additional slower channel only adds $1$ message every $10$ messages on the fast channel, a tolerable cost for high-end devices.
	

\fi
\vs{-0.8cm}
\section{Additional Evaluation Results}
\label{sec:additional_eval}
\vs{-0.4cm}

\begin{figure}[ht]
    \centering
    \includegraphics[width=0.8\columnwidth]{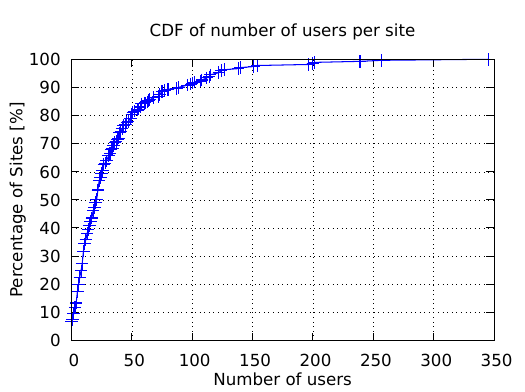}
    \caption{\small Distribution of users per ICRC site. $90\%$ of sites have less than $100$ users.}
    \label{fig:icrc-cdf}
    \vs{-0.5cm}
\end{figure}

\begin{figure}[ht]
	\hspace{-0em}
	\centering
	\includegraphics[width=0.8\columnwidth] {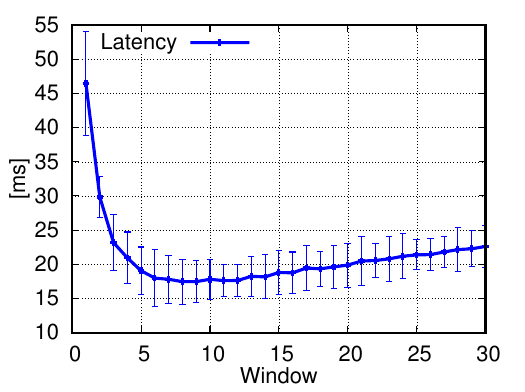}
	\caption{\small Effect of pipelining on latency. A window $W=7$ divides the latency by $2.25$ in comparison to the naïve $W=1$ approach.}
	\label{fig:window}
    \vs{-0.3cm}
\end{figure}

\begin{figure}[ht]
	\centering
	\includegraphics[width=0.8\columnwidth] {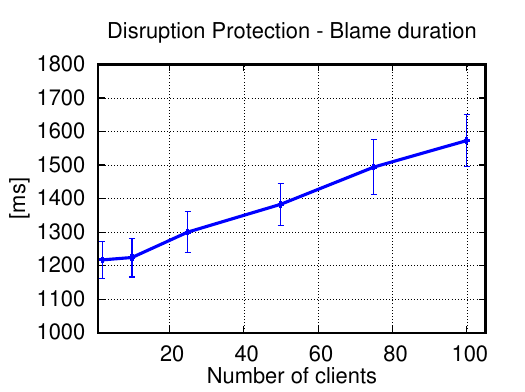}
	\caption{\small Duration of the blame procedure used to exclude a malicious client performing a disruption attack.
		Dissent's non-probabilistic version needs ``minutes to hours'' to exclude a disruptor~\citep{wolinsky2012dissent} (with probability $1$, unlike our protocol which has probability $1$ to detect but only $1/2$ to exclude).}
	\label{fig:disruption-blame}
    \vs{-0.3cm}
\end{figure}

\begin{figure}[ht!]
	\centering
	\includegraphics[width=0.8\columnwidth]{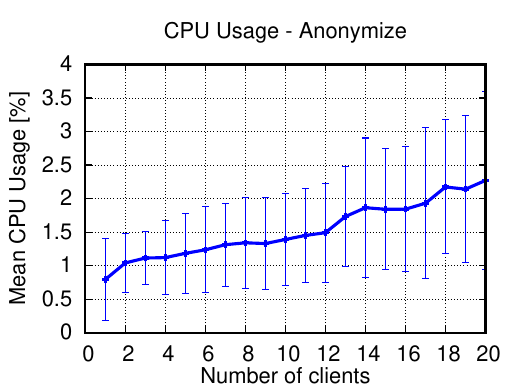}
	\caption{\small CPU usage on the relay during Anonymize, averaged over $10$ minutes.
		The client are real Android and iOS devices (hence the x axis stopping at $20$).
		The relay's hardware is a commodity server with a $3$GHz Xeon Dual Core and $2$GB of RAM\@.
		If we extrapolate the linear tendency, we estimate the mean CPU usage to be below $10\%$ with $100$ clients in this setup.}
	\label{fig:cpu}
\end{figure}

\begin{figure}[ht!]
\centering
\includegraphics[width=0.8\columnwidth]{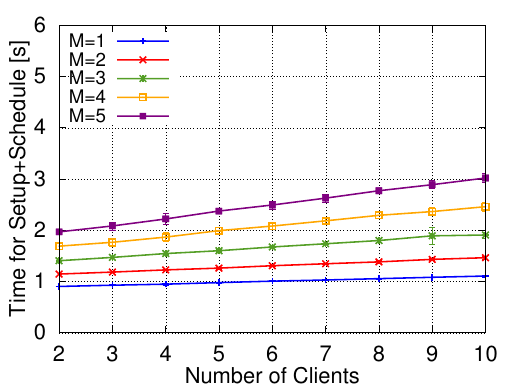}
\caption{\small Duration of \setup. This corresponds to the downtime in case of client churn.}
\label{fig:downtime}
\end{figure}

\para{CPU/Memory.} We briefly evaluate the CPU usage on the relay during an Anonymize round; our model estimates less than $10\%$ of average usage for $100$ clients on commodity hardware (Figure~\ref{fig:cpu}).
We tested the memory and CPU usage on an Android Nexus $5$X device; the device on which the measurements are done is doing light web browsing activities through \prifi.
During Anonymize, the mean CPU usage is light (below $5\%$), and the memory usage is moderate (stable at $50$~Mb; for comparison, Telegram uses $150$~Mb).
The energy consumption fluctuates between ``Light'' and ``Medium'' (estimated with Android Studio $3.2$ Beta) and is comparable to a VoIP call.
	
\end{document}